\documentclass[modern]{aastex63}

\usepackage{color}
\usepackage{url}

\shorttitle{Cosmological Evolution of FSRQs based on the {\it Swift}/BAT 105-month Catalog}
\begin{document}

\title{Cosmological Evolution of Flat-Spectrum Radio Quasars based on the {\it Swift}/BAT 105-month Catalog and Their Contribution to the Cosmic MeV Gamma-ray Background Radiation}

\author{Koyo Toda}%
\affiliation{Department of Physical Science, Hiroshima University, 1-3-1 Kagamiyama, Higashi-Hiroshima, Hiroshima 739-8526, Japan}%

\author{Yasushi Fukazawa}%
\email{fukazawa@astro.hiroshima-u.ac.jp}
\affiliation{Department of Physical Science, Hiroshima University, 1-3-1 Kagamiyama, Higashi-Hiroshima, Hiroshima 739-8526, Japan, fukazawa@astro.hiroshima-u.ac.jp}%
\affiliation{Hiroshima Astrophysical Science Center, Hiroshima University, 1-3-1 Kagamiyama, Higashi-Hiroshima, Hiroshima 739-8526, Japan}
\affiliation{Core Research for Energetic Universe (Core-U), Hiroshima University, 1-3-1 Kagamiyama, Higashi-Hiroshima, Hiroshima 739-8526, Japan}

\author[0000-0002-7272-1136]{Yoshiyuki Inoue}
\email{yoshiyuki.inoue@riken.jp}
\affiliation{Interdisciplinary Theoretical \& Mathematical Science Program (iTHEMS), RIKEN, 2-1 Hirosawa, Saitama 351-0198, Japan}
\affiliation{Kavli Institute for the Physics and Mathematics of the Universe (WPI), UTIAS, The University of Tokyo, Kashiwa, Chiba 277-8583, Japan}%Lines break automatically or can be forced with \\

\begin{abstract}
We present a new X-ray luminosity function of flat-spectrum radio quasars (FSRQs) utilizing the latest {\it Swift}/BAT 105-month X-ray source catalog. Contrary to previous studies of FSRQs in the X-ray band, using the luminosity-dependent  density evolution model, we find that FSRQs show evolutionary peaks at $z\sim1-2$ depending on luminosities. Our result is rather consistent with the evolution of FSRQs seen in the radio and GeV bands, although the number density is a factor of 5--10 smaller. We further explore the contribution of FSRQs to the cosmic MeV gamma-ray background radiation. We find that FSRQs can explain only $\sim3$\% of the observed MeV gamma-ray background fluxes around 1~MeV, indicating other populations are required. Future MeV gamma-ray observations will be keys for understanding the origin of the MeV gamma-ray background radiation.
\end{abstract}

\keywords{galaxies: active --- galaxies: jets --- X-rays: galaxies}
luminosity do not necessarily have the highest gamma-ray or radio

\section{Introduction}
After the discovery of the first quasar \citep{Schmidt1963}, to date, quasars are known to exist at least up to a redshift of $z=7.5$ corresponding to the first 690~million years of the cosmic time \citep{Banados2018}, and statistical samples now allows us to determine their cosmological evolution in a wide luminosity range up to $z\sim6$ \citep[e.g.,][]{Matsuoka2018}. 

The cosmological evolution of active galactic nuclei (AGNs) is a key for the understanding of the co-evolution between host galaxy and supermassive black hole (SMBH). Based on survey data in various wavelengths, the evolution of radio-quiet AGNs is well-studied \citep[see, e.g.,][for X-ray AGNs]{Ueda2014}, which have revealed a characteristic evolution history of SMBHs, so-called downsizing evolution. 

Among AGN populations, blazars whose relativistic jets point to us show another aspect of SMBHs. Their emission spans from radio to gamma-ray, and it is dominated by non-thermal jet emission. Theoretically, it is discussed that jets are launched through the Blandford-Znajek mechanism \citep{Blandford1977}. Observations of jetted AGNs suggest that jet power correlates with accretion rate \citep[e.g.,][]{Ghisellini2014,Inoue2017} and spin parameter would determine the radio-loudness distribution \citep[e.g.,][]{Sikora2007}. Therefore, the cosmological evolution of blazars would shed light not only on the evolution of SMBHs themselves but also on the cosmic history of jet activity in AGNs. 

Since blazars are rare in the sky, the construction of blazar samples requires an all-sky survey data. Therefore, all-sky surveys such as {\it Fermi} Large Area Telescope \citep[LAT,][]{Atwood2009} and {\it Swift} Burst Alert Telescope \citep[BAT,][]{Gehrels2004} provide unique opportunities for the study of blazar evolution. Utilizing blazar survey data, cosmological evolution of blazars have been extensively studied in literature \citep[e.g.,][]{Padovani1993,Narumoto2006,Inoue2009,Ajello2009,Ajello2015}. 

Blazars are classified into BL Lac objects and flat-spectrum radio quasars (FSRQ), based on their optical line properties \citep{Urry1995}. BL Lac objects often lack optical emission lines \citep[see, e.g.,][]{Shaw2013}, while FSRQs show strong optical lines. FSRQs are also more luminous than BL Lacs \citep{Fossati1998, Kubo1998, Ghisellini2017}, and they show a hard spectrum in the X-ray band and peaking in the MeV band \citep[e.g.,][]{Blom1995}. Therefore, redshift complete sample is available for FSRQs, and their X-ray luminosity function (XLF) allows us to investigate the evolution of jetted AGNs in the early universe.

The currently available FSRQ XLF is constructed with 26 FSRQs based on {\it Swift}/BAT 22-month survey data \citep[][hereinafter A09]{Ajello2009}. FSRQ evolution in the GeV band is recently updated using 186~FSRQ samples detected by {\it Fermi}/LAT \citep{Ajello2012, Ajello2015}. Since X-ray and gamma-ray are believed to be generated by the same emission mechanism in FSRQs, inverse Compton scattering, it is naturally expected that X-ray and gamma-ray evolutions show a similar tendency. However, their cosmological evolution is known to be different. For luminous FSRQs, the evolutionary peak is reported as $z\sim4.3$ in the X-ray band (A09), while it is $z\sim2$ in the gamma-ray band \citep{Ajello2012}. This contradiction would be due to the small sample size of X-ray FSRQ and different assumed luminosity function form between X-ray and gamma-ray band. 

As FSRQs have spectral peaks in the MeV band, FSRQs are argued as a possible origin of the cosmic MeV gamma-ray background radiation (A09), which is still an intriguing mystery. However, other candidates are also suggested such as non-thermal coronal emission from radio-quiet AGNs \citep[e.g.,][]{Inoue2008, Inoue2019} and dark matter particles \citep[e.g.,][]{Olive1985,Ahn2005_DM1,Ahn2005_DM2}. A detailed understanding of the FSRQ XLF is important to quantitatively understand FSRQs' contribution to the MeV background radiation.

Recently, {\it Swift}/BAT released their latest all-sky catalog data using their 105-month survey data \citep{Oh2018}. In this paper, we report the evolution of FSRQs utilizing this latest {\it Swift}/BAT survey catalog. We further compare our result with available FSRQ evolution in the GeV and radio band. Moreover, we study the contribution of FSRQs to the cosmic MeV gamma-ray background radiation. Throughout the paper, we assume cosmological parameters of $\Omega_{\lambda}=0.7$,  $\Omega_{M}=0.3$ and $H_0$=70 km s$^{-1}$ Mpc$^{-1}$.

\section{Selection of FSRQ sample}

We selected a sample of X-ray FSRQs from the {\it Swift}/BAT 105~month catalog, referring to the 5th BZCAT catalog\footnote{\url{https://heasarc.gsfc.nasa.gov/w3browse/all/romabzcat.html}}. The 5th BZCAT catalog contains 3561 blazar sources, including 1909 FSRQ (referred to as 5BSQ in this catalog), based on multi-wavelength information \citep{Massaro2015}.

Since the BZCAT catalog is not complete along the Galactic plane region, we restricted sources at a Galactic latitude of $\left|b\right|>15^{\circ}$. Positional error of BAT 105 month sources is $\{(30.5/{\rm SNR})^2+0.1^2\}^{1/2}$~arcmin at the 90\% confidence level \citep{Oh2018}, where ${\rm SNR}$ is a signal-to-noise ratio. When we adopt this value as a criterion for positional coincidence between the BAT catalog and the BZCAT FSRQ catalog, the number of samples becomes less than 20, which is smaller than the sample size of A09. Therefore, as a secure criterion, we set the matching radius as 10~arcmin. A chance probability that a non-counterpart BZCAT blazar matches a BAT source is small enough $\sim0.3$\%. When matching objects, we also consider redshift information in both catalogs. We only choose the BZCAT sources whose redshift differences from the BAT sources is $<0.01$ for objects with a redshift $z$ of $<2$ and $<0.06$ for $z\ge2$. This is because of a larger redshift measurement error of distant objects.

We selected the FSRQ sample where all the above thresholds are satisfied, and as a result, 53 sources are selected as listed in Table \ref{sample}. The most distant one is located at $z=4.715$ (SWIFT~J1430.6+4211). Compared to the A09 sample, the sample number increases by a factor of $\sim2$. Our sample includes most of A09 sample FSRQs except three objects. Two of the excluded objects had a low SNR around 5 in A09, while the other one had an SNR of $\sim$10. One possibility is due to time variability; a bright phase was detected in A09, while it is not detected in the 105-month catalog after the flux is averaged over 105~months.

\begin{figure*}[t]
\includegraphics[width=.5\textwidth]{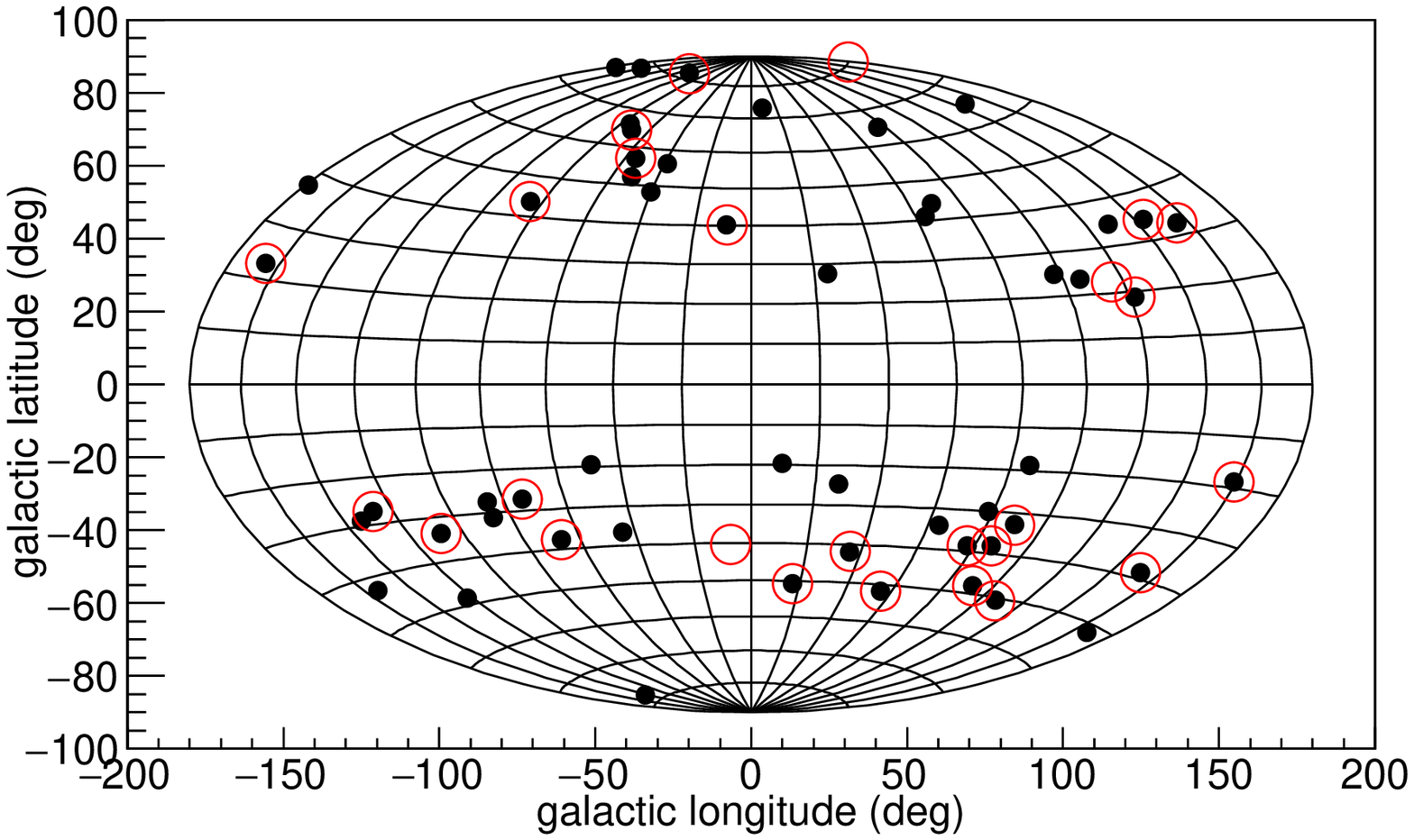}\hfill
\includegraphics[width=.5\textwidth]{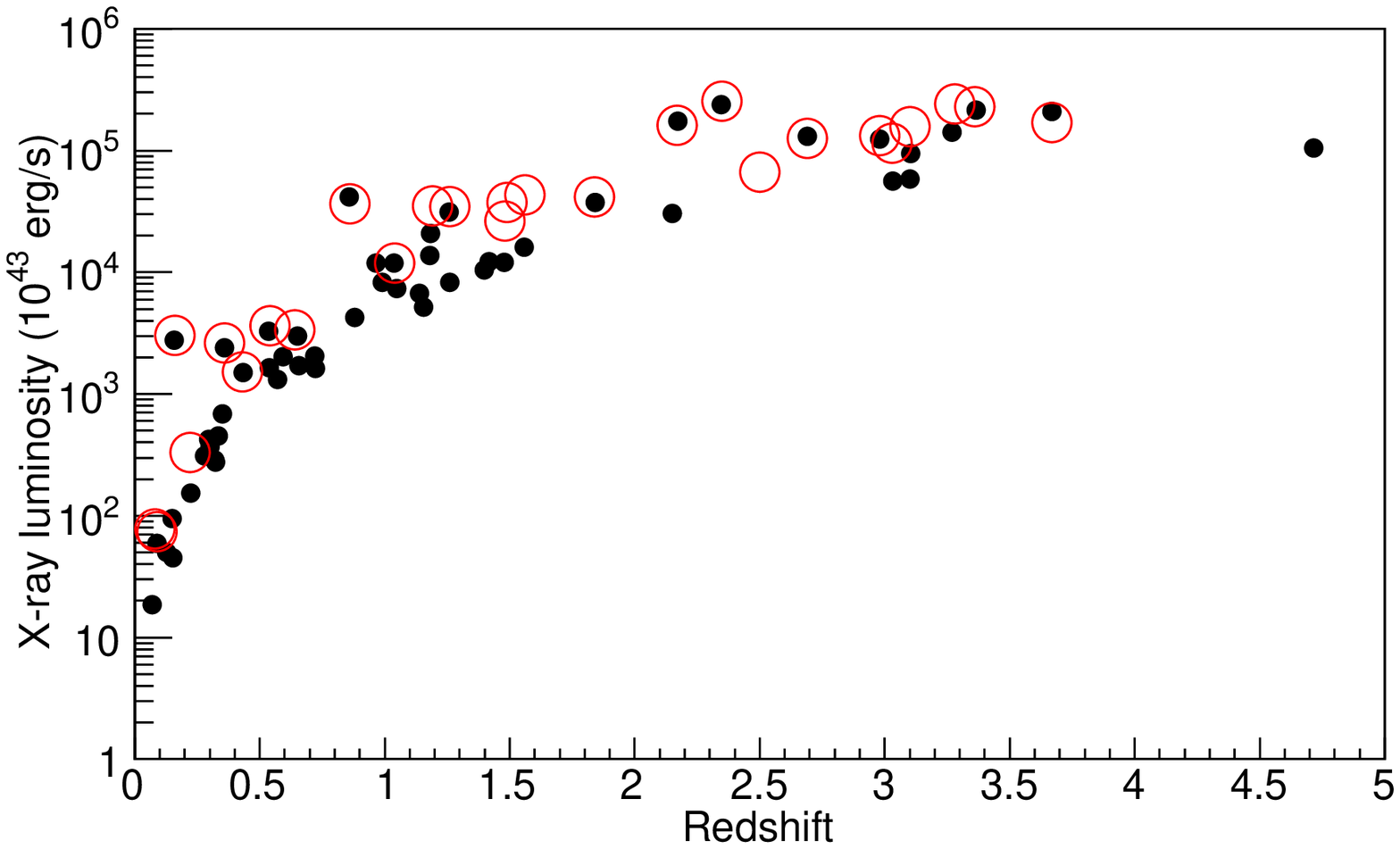}
\\[\smallskipamount]
\includegraphics[width=.5\textwidth]{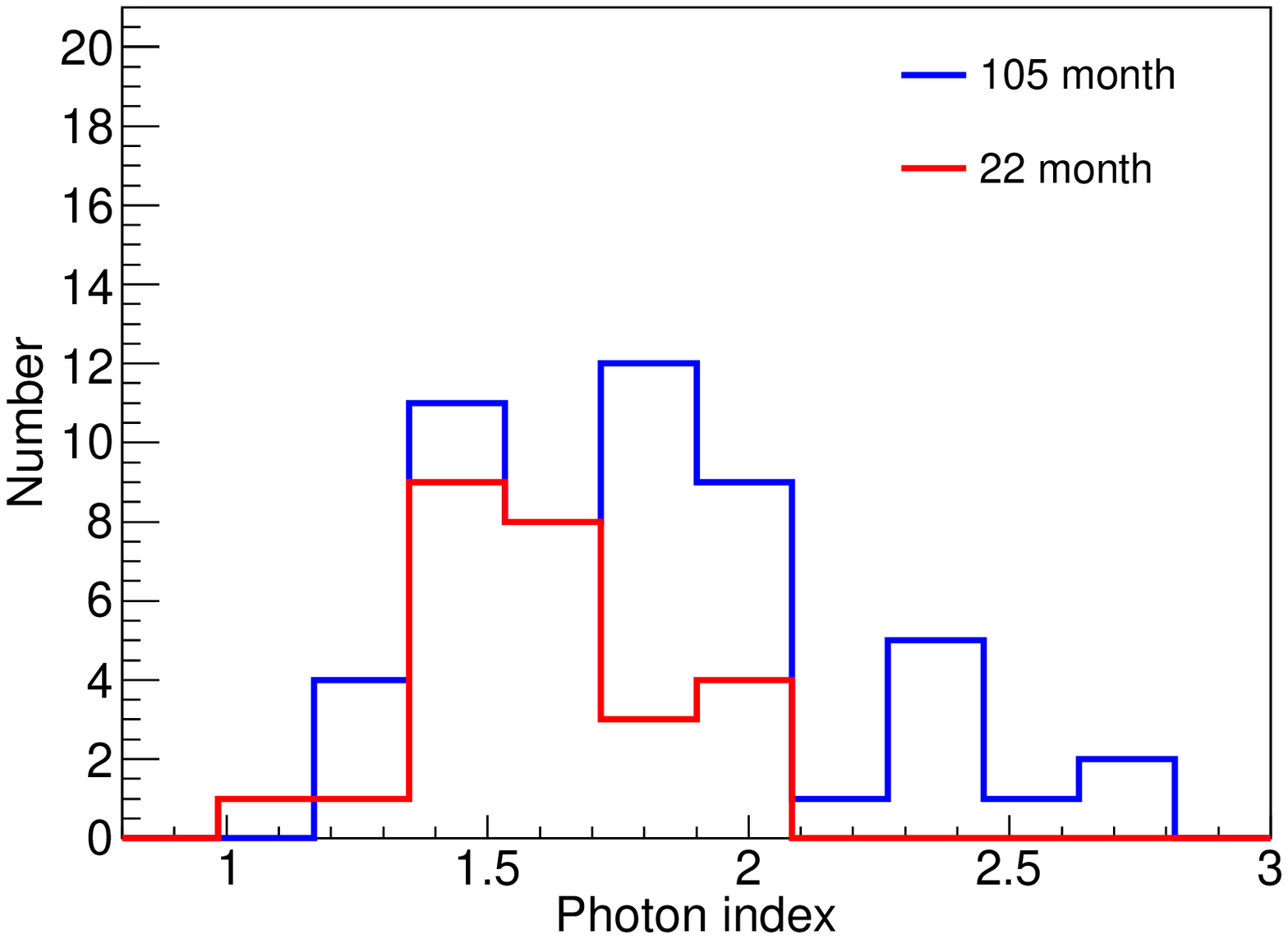}\hfill
\includegraphics[width=.5\textwidth]{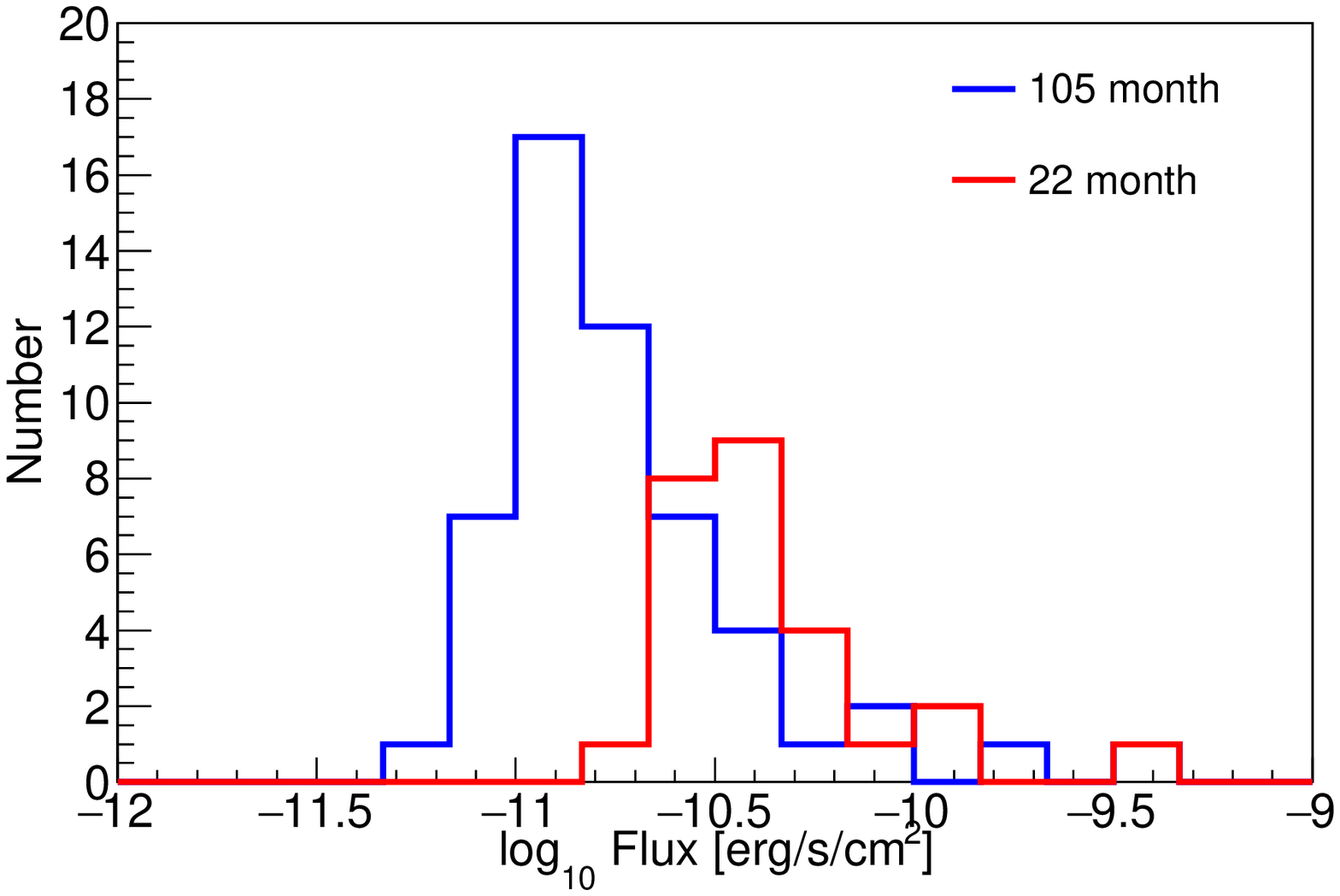}\\[\smallskipamount]
%\caption{Properties of our sample BAT FSRQs. {\it Left-top}: Redshift
%distribution, {\it right-top} BAT X-ray flux (14--195 keV) distribution, {\it left-bottom} BAT photon %index distribution, and {\it right-bottom} luminosity distribution (14-195 keV). Blue and red solid line %represent FSRQs in our sample and the A09 sample, respectively.}
\caption{Properties of our sample BAT FSRQs. {\it Left-top}: Sky position in the Galactic coordinate, {\it right-top}: redshift vs BAT X-ray luminosity distribution, {\it left-bottom}: BAT photon index distribution and {\it right-bottom}: BAT X-ray flux (14--195 keV) distribution. For the top two panels, black filled circles and red open circles represent FSRQs in our sample and the A09 sample, respectively.   For the bottom two panels, blue and red solid line represent FSRQs in our sample and the A09 sample, respectively.}
\label{comp22}
\end{figure*}

Figure. \ref{comp22} shows comparisons of FSRQ properties between A09 and our sample in sky position, redshift vs luminosity, photon index, and flux. The flux distribution of our sample shifts to lower fluxes comparing to A09, because the 105-month catalog contains new fainter FSRQs and the flux of most of FSRQs in the 22-month catalog becomes lower in the 105-month catalog. The latter cause is possibly due to time variability. While the number of distant FSRQs does not significantly increase, low-z ($<1.5$) and medium luminosity sources increase. Also, FSRQs with a larger photon index (i.e., a softer X-ray spectrum) with $>1.9$ are contained in our sample. These imply that A09 might have a selection bias so that FSRQs with a low-z mid-luminosity or a softer X-ray spectrum are lacked.

\section{Luminosity Function}

Following recent studies of X-ray and gamma-ray luminosity functions
of blazars \citep{Narumoto2006,Inoue2009,Ajello2009,Ajello2012}, we adopt a luminosity function of the luminosity-dependent density evolution (LDDE) model,

\begin{eqnarray}
  \Phi(L_X,z)&=&\frac{d^2N}{dzdL_X}\\ \nonumber
  &=&\frac{A}{\ln(10)L_X}
  \left[\left(\frac{L_X}{L_*}\right)^{\gamma_1} +
   \left(\frac{L_X}{L_*}\right)^{\gamma_2}  \right]^{-1}\\ 
   &\times&
  \left[\left(\frac{1+z}{1+z_c(L_X)}\right)^{p_1} +
   \left(\frac{1+z}{1+z_c(L_X)}\right)^{p_2}    \right]^{-1} ,
\end{eqnarray}

where $z_c(L_X)=z_c({L_X}/{10^{47.5}})^{\alpha}$. Here, $z$ and $L_X$ are a redshift and an X-ray luminosity in the 14--195 keV band. Parameters to be constrained are a normalization $A$, a characteristic X-ray luminosity $L_*$, a characteristic redshift $z_c$, two luminosity indices $\gamma_1$ and $\gamma_2$, two redshift indices $p_1$, $p_2$, and $\alpha$.

\begin{table}
\begin{center}
\caption{Best-fit parameters for the LDDE model}
%\hspace{-1.8cm}
\begin{tabular}{cccccc}
\hline
\hline
$\log_{10} A^a$ & $\gamma_2$ & $p_1$ & $p_2$ & $z_c$ & $\alpha$ \\ 
\hline
$-13.02_{-0.35}^{+0.25}$ & $0.80_{-0.08}^{+0.05}$ & $3.58_{-3.08}^{+0.42}$ & $-7.7_{-0.3}^{+1.5}$ & $1.36_{-1.03}^{+0.64}$ & $0.42_{-0.06}^{+0.18}$ \\
\hline
\end{tabular}
\label{bestfit}
\end{center}
$a$: In the unit of Mpc$^{-3}$.\\
Errors are in 1-$\sigma$ uncertainty region.
\end{table}

\begin{figure*}[t]
\gridline{
\includegraphics[clip,width=.33333\textwidth]{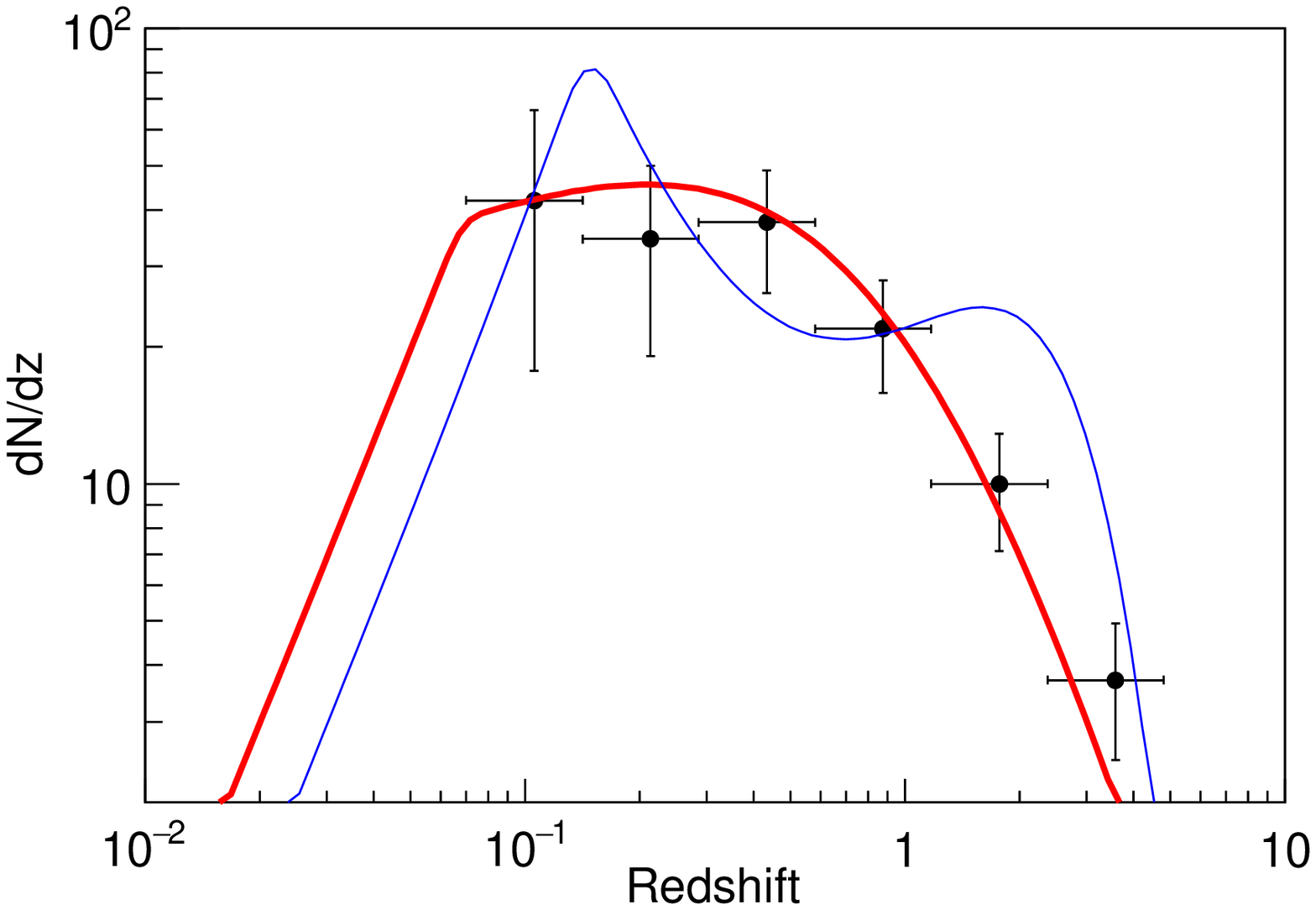}
\includegraphics[clip,width=.33333\textwidth]{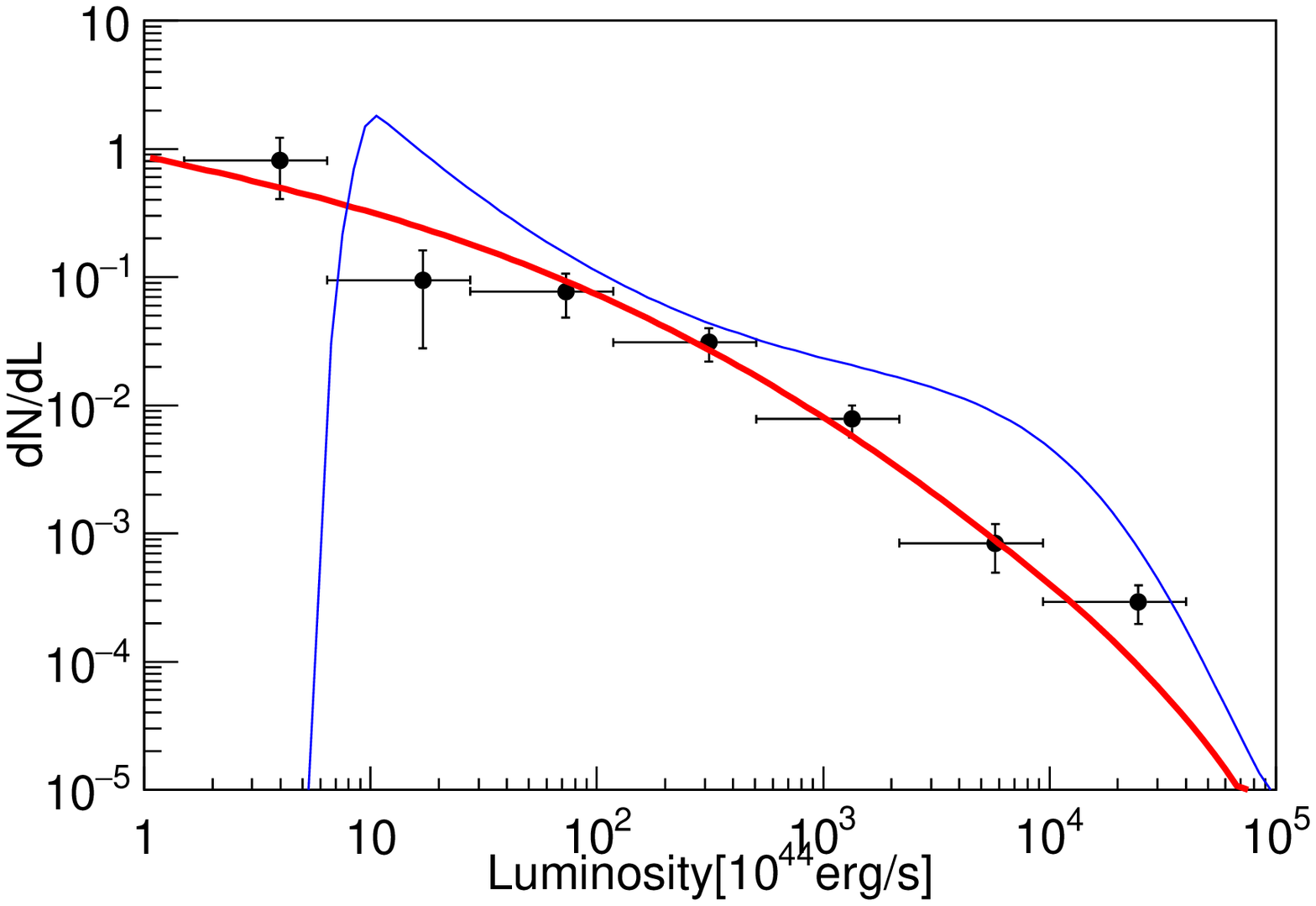}
\includegraphics[clip,width=.33333\textwidth]{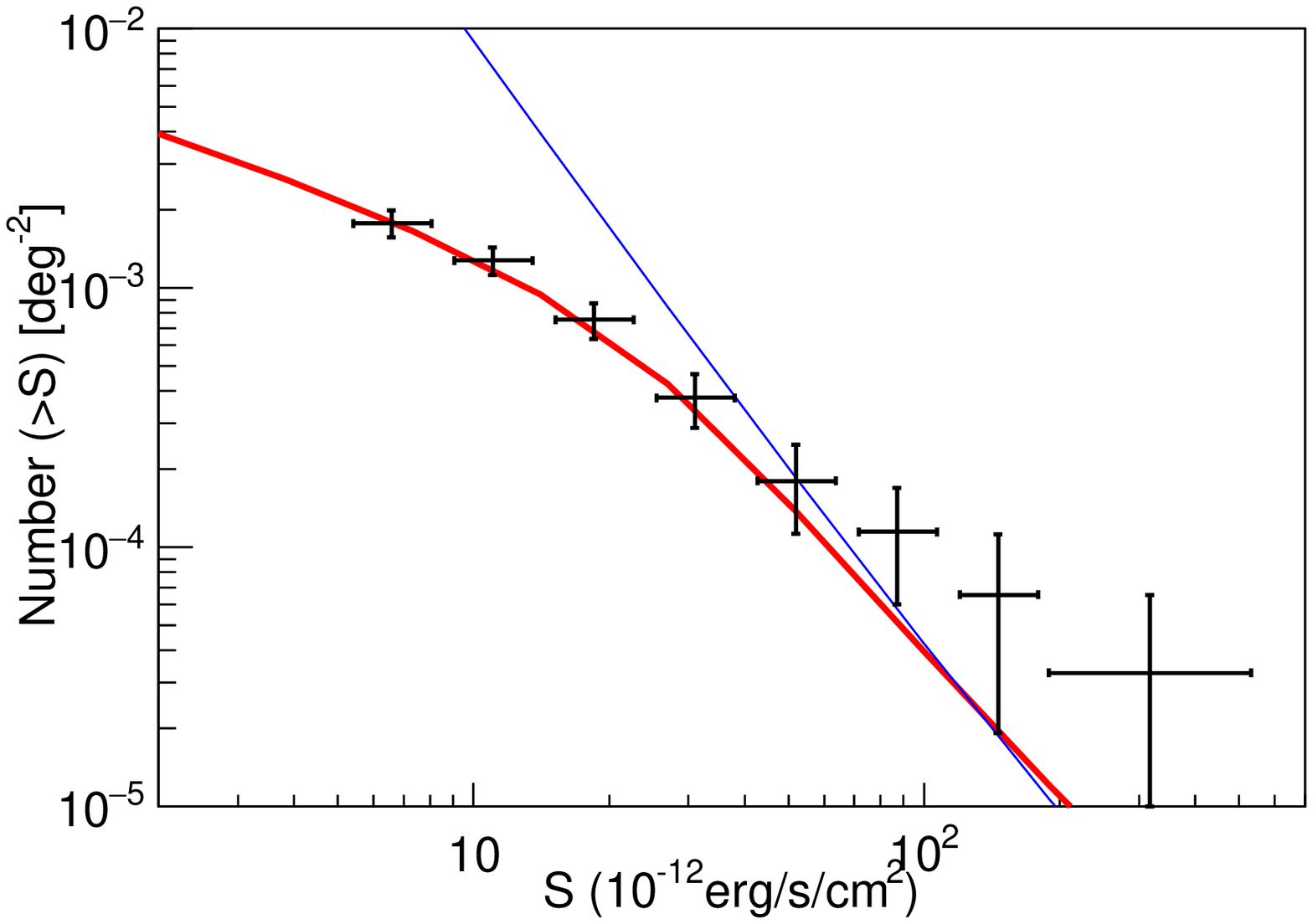}
}
\caption{{\it Left}: Redshift distribution of our sample, prediction from the best-fit LDDE model (red) and that from the A09 model 10 (blue). {\it Middle} and {\it Right} are that for luminosity and cumulative source number count, respectively.}
\label{zldist}
\end{figure*}

Since our sample contains 53 sources only, all parameters cannot be constrained simultaneously. We found that two indices on luminosity dependence become a similar value around 0.8 when we make both of them free in our fits, and thus hereafter, we fix the index of higher luminosity side to 5.0, and the characteristic luminosity is fixed to $10^{51}$ erg s$^{-1}$. This corresponds to a single-power-law luminosity dependence without a break up to the characteristic luminosity. 

We perform maximum likelihood analysis to determine LDDE model parameters. Likelihood function $\mathcal{L}$ is defined as follows:
\begin{eqnarray} \nonumber
 \ln \mathcal{L} &=& \sum^{N_{\rm obs}}_i \log \left( \Phi (L_{Xi},z_i)  \right)\\
   &-&  \int_{z_{\rm min}}^{z_{\rm max}} dz
  \int_{L_{X, \rm min}}^{L_{X, \rm max}} dL \Phi(L_X,z)S(L_X,z),
\label{Lfanction}
\end{eqnarray}

where $S(L_X,z)$ is the sky coverage of the BAT 105 month catalog and we adopt the right panel in the figure 10 in \citet{Oh2018}. We set $z_{\rm min}=0$, $z_{\rm max}=6$, $L_{X, \rm min}=8\times10^{43}$ erg s$^{-1}$, and $L_{X, \rm max}=10^{50}$ erg s$^{-1}$. The redshift and luminosity ranges are based on the observed ranges (See Fig. \ref{comp22}). Best-fit parameters are determined by finding the maximum likelihood $\mathcal{L}_{\rm max}$, and 1-$\sigma$ errors are estimated as a parameter range that satisfies $2\left(\log \mathcal{L}-\log \mathcal{L}_{\rm max}\right)>-1$ (A09) when the other parameters are left free. Table. \ref{bestfit} summarizes the best-fit parameters of LDDE.

Figure. \ref{zldist} shows the redshift, luminosity, and source count distributions for data and the best-fit LDDE  model. Those for the model are calculated as
\begin{eqnarray}
 \frac{dN}{dz} &=& 4\pi \int_{L_{\rm min}}^{L_{\rm max}} \Phi(L_X,z)S(L_X,z) dL_{X} \frac{dV}{dzd\Omega}\\
\frac{dN}{dL_X} &=& 4\pi
 \int_{z_{\rm min}}^{z_{\rm max}} \Phi(L_X,z)S(L_X,z)dz \frac{dV}{dzd\Omega} \\
 N(>S_0)&=&\int^{z_{\rm max}}_{z_{\rm min}} dz \frac{dV}{dzd\Omega}
  \int^{\infty}_{L_0}  \Phi(L_{X},z)dL_{X}
\label{dNdzdNdL}
\end{eqnarray}
where ${dV}/{dzd\Omega}$ is a comoving volume per redshift per solid angle, $L_0 = 4\pi D_L(z)^2 S_0 (1+z)^{\Gamma -2}$. $S_0$ is the observed X-ray flux and $\Gamma$ is a power-law photon index of FSRQ X-ray spectra. Here, we fix it to 1.807, a mean value of our BAT FSRQ sample. For the logN-logS relation, we correct data points by the sky coverage factor. As expected, the best-fit LDDE model nicely reproduces these distributions well, compared to the A09 best-fit model (model 10 in A09), especially for low flux regime where the A09 model could not trace.

\begin{figure*}
\includegraphics[width=.5\textwidth]{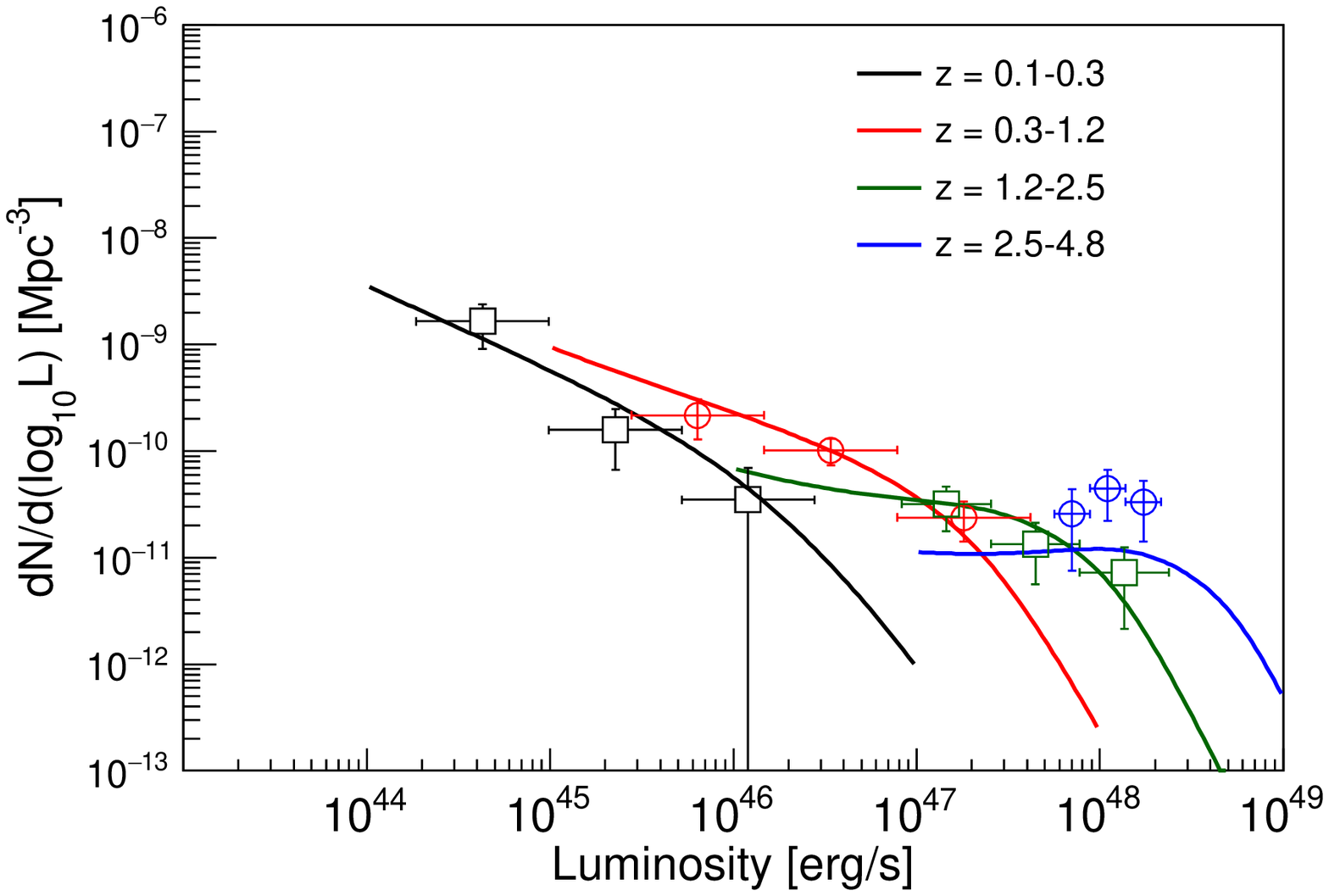}
\hfill
\includegraphics[width=.5\textwidth]{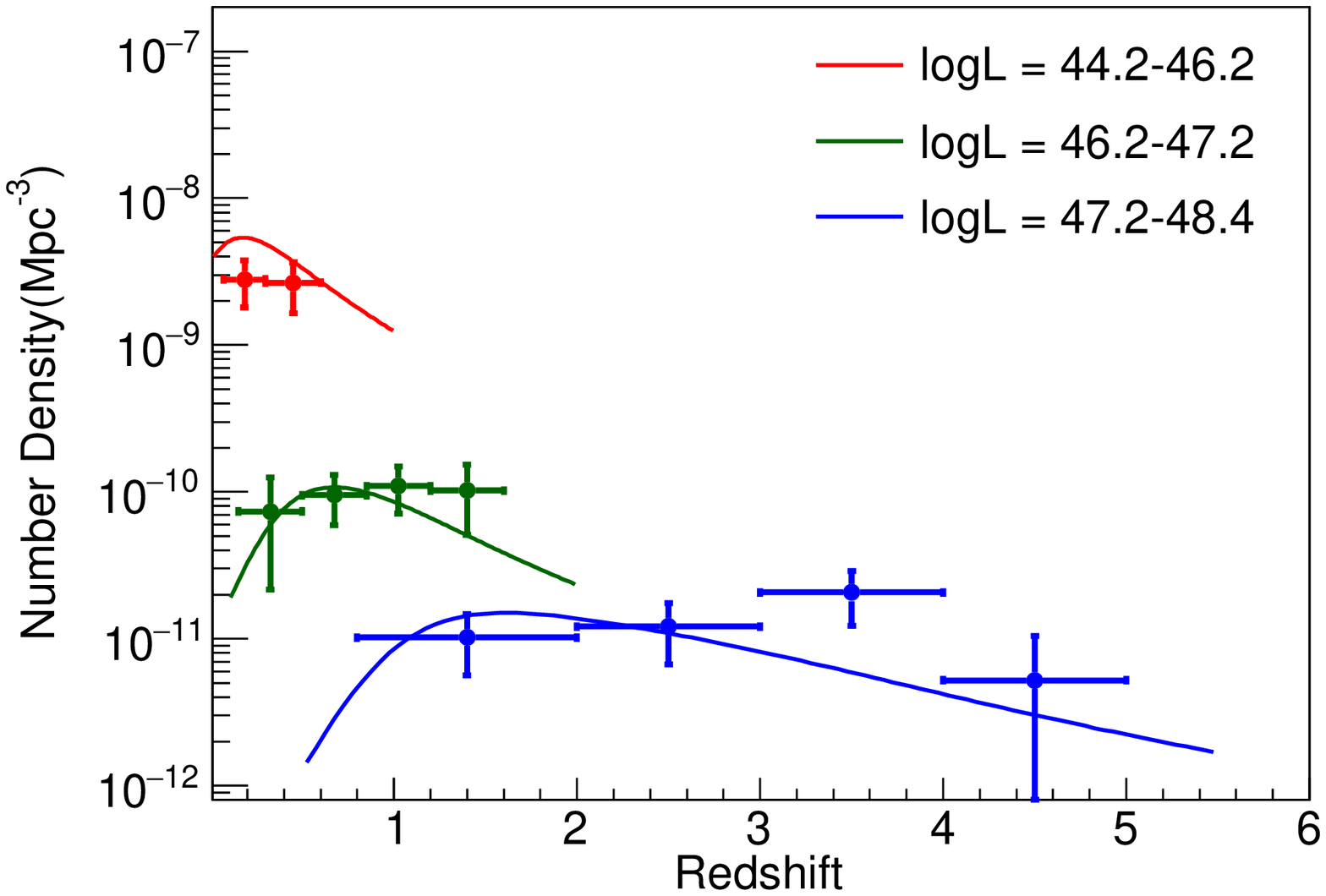}
\caption{{\it Left}: X-ray luminosity function of FSRQs in various redshift bins. Model curves correspond to the best-fit LDDE model at different redshift bins. Model curves are multiplied by $N^{\rm obs}/N^{\rm mdl}$. {\it Right}: Same as the {\it Left}, but for the comoving number density of FSRQs in various luminosity bins. Data points are deconvolved by dividing them by $N^{\rm obs}/N^{\rm mdl}$.}
\label{LF}
\end{figure*}

Figure. \ref{LF} shows a visual representation of the best-fit luminosity function (left) and comoving number density (right). For the left panel of figure \ref{LF}, we plot the model curves by the ``$N^{\rm obs}/N^{\rm mdl}$'' method \citep{laFranca1997,Miyaji2001}; we multiply the luminosity function $\Phi(L_{X,i},z_i)$ of the $i$-th bin of a luminosity $L_{X,i}$ and a redshift $z_i$  by $N_i^{\rm obs}/N_i^{\rm mdl}$ where $N_i^{\rm obs}$ and $N_i^{\rm mdl}$ are the observed and the predicted number of FSRQ in that bin. For the right panel, we deconvolve the observed data points by dividing them by $N_i^{\rm obs}/N_i^{\rm mdl}$.

The best-fit LDDE model reproduces these data well. The so-called downsizing evolution is clearly seen. The peak redshift in the highest-luminosity bin is around $z\sim1.8$, similar to the radio and GeV gamma-ray ones of $z$=1.8--2.2 \citep{Ajello2012,Mao2017}. At $z\sim3$--4, the model prediction is lower than the data. We note, however, that this discrepancy is within the data error bar and model uncertainty (discussed later). Other parameters are also similar to those of the GeV and radio band.

%The best-fit LDDE model reproduces these data well, except for one bin with a redshift of 3--4. As discussed later, the model curve has uncertainty. A high redshift evolution indicator $p_1$ becomes around 3.6, somewhat smaller but similar to other cases of 4--6. The peak redshift in the highest-luminosity bin is around $z\sim1.8$, similar to the radio and GeV gamma-ray ones of $z$=1.8--2.2. A luminosity-dependence index of peak redshift is somewhat large around 0.4, in comparison with radio and GeV cases of 0.1--0.3. A luminosity-dependence slope $\gamma_1$ and a  low-z redshift-dependence slope $p_2$ are similar to those of other cases.

\section{Cosmic X-ray and MeV gamma-ray background radiation}
The FSRQ contribution to the cosmic X-ray background (CXB) radiation is calculated as
\begin{eqnarray}\nonumber
 F_{\rm CXB}(E_0) &=& \int_{z_{\rm min}}^{z_{\rm max}}
  \frac{dV}{dzd\Omega} dz
  \int_{L_{\rm min}}^{L_{\rm max}} dL_{X}\\
  &\times&\Phi(z,L_{X})  F(z,L_{X},E_0),
\label{CXB}
\end{eqnarray}
where $F(z,L_X,E_0)$ is a flux at an energy of $E_0$ in the observer frame for a source with a redshift $z$ and a X-ray luminosity $L_X$ (14--195 keV). We have assumed two spectral shapes for FSRQ spectra; a single power-law and a broken power-law. A photon index of the single power-law and a low-energy photon index of the broken power-law is set to 1.807 (Fig.~\ref{comp22}), which is the mean BAT photon index over our sample of FSRQs. A high-energy photon index of the broken power-law is set as 2.5 as A09. 

\begin{figure*}
\includegraphics[width=.5\textwidth]{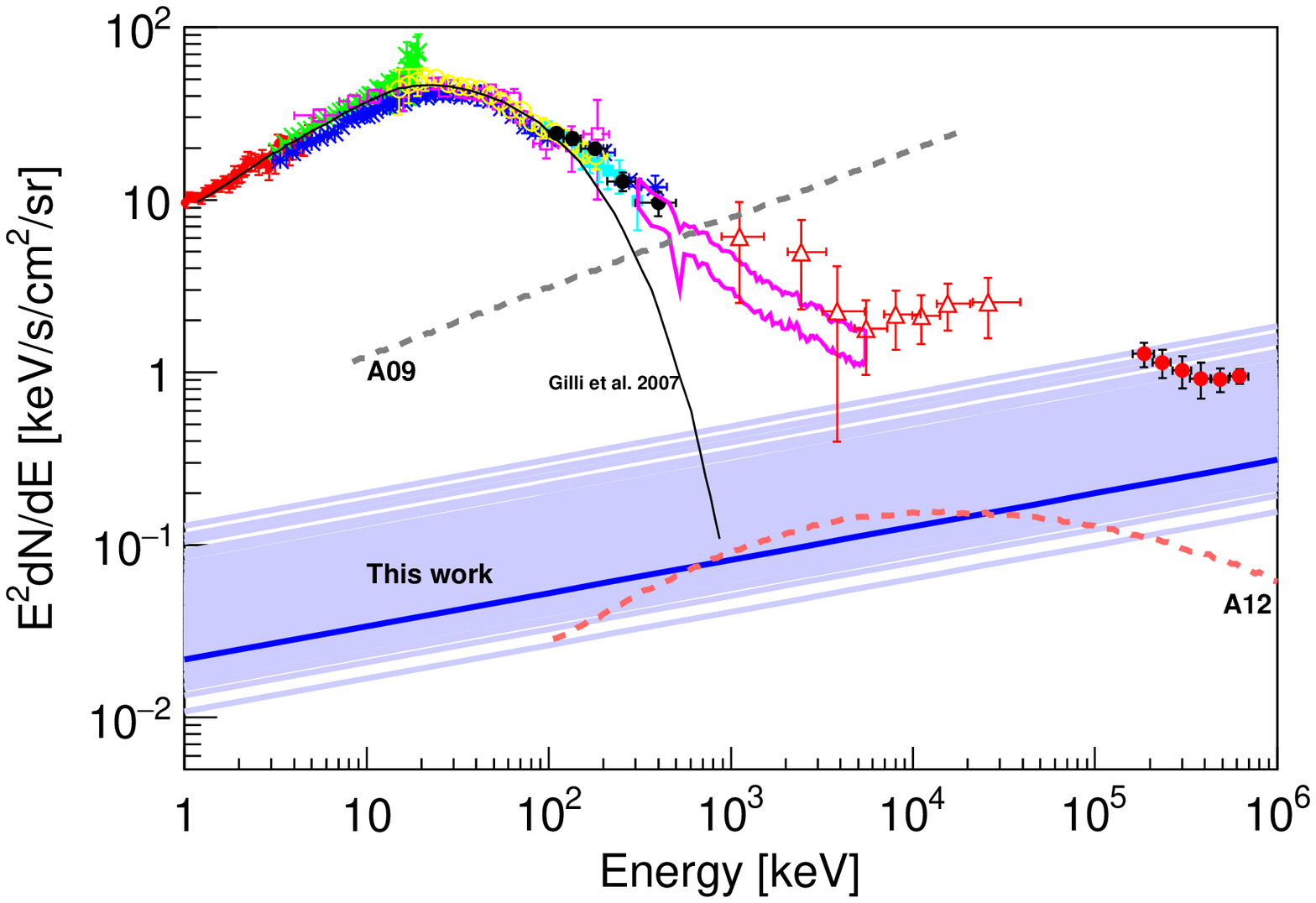}\hfill
\includegraphics[width=.5\textwidth]{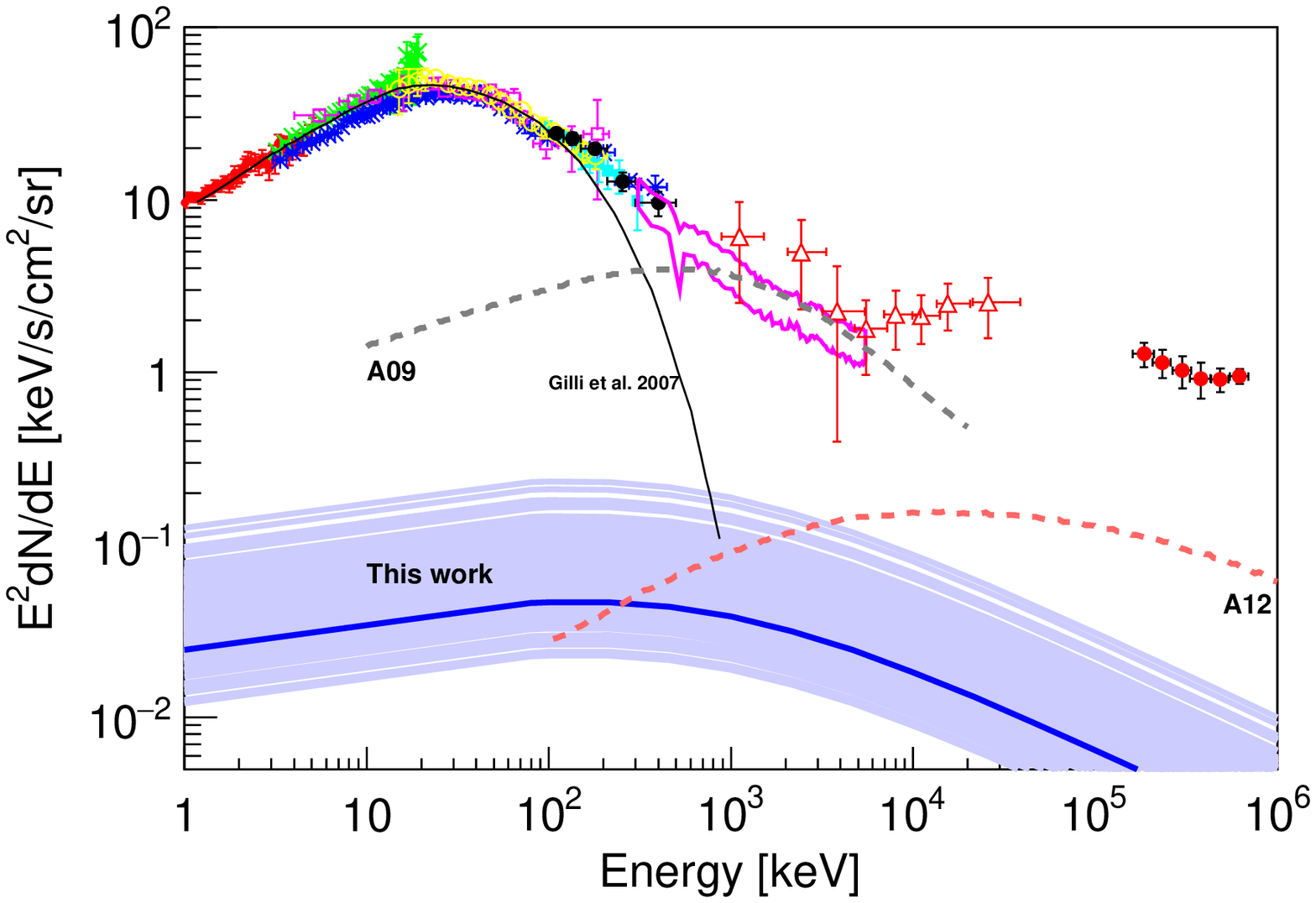}\hfill
\caption{Contribution of FSRQ to the cosmic X-ray and MeV gamma-ray background radiation, estimated by the best-fit LDDE model. Left and right panel shows a contribution by assuming a single power-law and a broken power-law, respectively, for FSRQ spectra. Blue solid line and thin band represent an estimation and its error region, respectively, from our best-fit model. Black and red dashed-line represents an estimation in A09 and \citet{Ajello2012}, respectively. Black thin line represents an estimation of \citet{Gilli2007}. Red, light blue, yellow, open squares are CXB measuments by INTEGRAL \citep{churazov2007}, HEAO-1 A4 \citep{Kinzer1997}, Swift/BAT \citep{Ajello2008},  respectively. Red open triangle and black filled squares are those by COMPTEL \citep{Weidenspointner2000} and Nagoya balloon \citep{Fukada1975}, respectively. Cyan solid line is that by SMM \citep{Watanabe1997}. Red, light green, blue crosses are those by ASCA \citep{Gendreau1995}, XTE \citep{Revnivtsev2003}, HEAO-1 A2 \citep{Gruber1999}, respectively. Red filled circles are those by Fermi/LAT \citep{Ackermann2015}.}
\label{cxb}
\end{figure*}

The left panel of figure \ref{cxb} shows the model calculation results (blue solid line), together with various measurements of the cosmic X-ray and gamma-ray background radiation. We assume the case with a single power-law spectrum for FSRQs. The observed data are the same as A09, however we added the {\it Fermi}/LAT measurement \citep{Ackermann2015}. We also plot the predicted contribution from Seyfert galaxies \citep{Gilli2007} (black thin line) and FSRQs (black and red dashed-line); the latter two are based on the BAT 22-month catalog (A09) and {\it Fermi}/LAT \citep{Ajello2012}, respectively. Error region of our model prediction is also displayed (blue thin band). We evaluate the error region of our model by varying each of fitted LDDE parameters randomly within a 1 $\sigma$ error by considering parameter correlations with the correlation matrix. 

As clearly seen from the figure, it would be difficult for FSRQs to explain the whole cosmic MeV gamma-ray background fluxes. They can explain only $\sim3$\% of the MeV gamma-ray background around 1~MeV. Our prediction on the background flux is an order of magnitude lower in the hard X-ray band than that by A09, likely due to the difference of FSRQ luminosity function. Higher prediction of FSRQ contribution by A09 is due to undersampling of low redshift fSRQs as described in \S~2, leading to relatively abundant high redshift FSRQs among their sample.

In the right panel of figure \ref{cxb}, we plot the same as in the left panel, but plot the prediction by assuming that FSRQ spectra are broken power-law, as presented by A09. The break of the spectrum was required so that the prediction does not exceed the measurement above 1 MeV in A09. However, we note that the break energy of 1 MeV is lower than that of the FSRQ gamma-ray SED model in \citet{Ajello2012}, where the break is around 100 MeV as shown in the figure. If we adopt the SED model of \citet{Ajello2012}, the expected FSRQ contribution based on our XLF will become ten times higher than the \citet{Ajello2012} curve in the entire energy range, where we assumes that the FSRQ population in the hard X-ray band is the same as that of the GeV band. Therefore, the missing LAT counterpart of BAT FSRQs indicates the hidden population of steep and faint GeV gamma-ray FSRQ, suggesting that the break energy is less than 100 MeV.

\section{Discussion}
\subsection{Comparison with Previous Works}

\begin{figure}
\includegraphics[width=.5\textwidth]{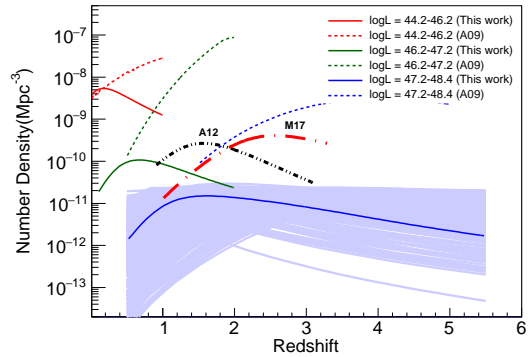}
\caption{Comparison of the evolution of FSRQs based on our result with the previous results of 22-month {\it Swift}/BAT hard X-ray (A09), {\it Fermi}/LAT GeV gamma-ray \citep[][A12]{Ajello2012}, and radio band \citep[][M17]{Mao2017}. Model curves are not multiplied by $N_i^{\rm obs}/N_i^{\rm mdl}$. Solid lines are our results, dot lines are those of BAT 22 month (A09), dot-dashed and dot-dot-dashed lines are those of radio (M17) and {\it Fermi}/LAT (A12), respectively. Red, black, and blue curves correspond to the luminosity bin of $10^{44.2-46.2}$ erg s$^{-1}$, $10^{46.2-47.2}$ erg s$^{-1}$, and $10^{47.2-48.4}$ erg s$^{-1}$, respectively.Thin blue band represents an error region of the model at the highest luminosity bin for our result. Results of the BAT 22-month catalog are plotted after correcting the different energy bands between our works and A09. For the {\it Fermi}/LAT and radio ones, curves of the highest luminosity bins in the original papers are plotted. Redshift ranges are restricted to those in the original references. 
}
\label{LFcmp}
\end{figure}

Figure. \ref{LFcmp} shows the comoving density of FSRQs based on our results, together with the previous results of the 22~month {\it Swift}/BAT hard X-ray (A09), {\it Fermi}/LAT GeV gamma-ray \citep{Ajello2012}, and radio band \citep{Mao2017}. To compare the models, in this figure, we do not multiply model curves by $N_i^{\rm obs}/N_i^{\rm mdl}$. The function of A09 is based on their modified pure luminosity evolution model with a double power-law form (model 10 in their paper), while others are on the LDDE. We also show the error region (thin blue band) of our model at the highest-luminosity end by varying each of fitted LDDE parameters randomly within a 1 $\sigma$ error by considering parameter correlations with the correlation matrix. Note that model curves in the original papers \citep{Ajello2012, Mao2017} are normalized by luminosity,  and calculated at a representative luminosity in each curve, in a different way from here. 

Our results are quite different from those of A09, even though both works study the XLF of FSRQs. This difference is likely due to the following two reasons. First, we use the 105~month {\it Swift}/BAT survey data having 53 FSRQs upto $z=4.715$, while A09 used the 22~month {\it Swift}/BAT survey data having 26 FSRQs upto $z=3.67$. As a result, observed redshift distributions are quite different. We have abundant low redshift FSRQs comparing to A09 (See Fig.~\ref{comp22}). Second, we adopt the LDDE model for the XLF formulation, which is known to well reproduce the AGN evolutions for radio-quiet AGNs \citep[e.g.,][]{Ueda2014} and blazars \citep[e.g.,][]{Inoue2009,Ajello2015}, while A09 utilized the modified pure luminosity function formula. Likely because of this formulation, A09 predict the evolutionary peak at $z\sim4.3$ which is out of the range of their FSRQ sample data.

Our curve at the highest luminosity bin has a different number density comparing to the GeV and radio ones, regardless of the same function form of the luminosity function. Radio and GeV FSRQs in the highest luminosity bins seem richer than X-ray one by a factor of 5--10 at a redshift of 1--3. The orders of the luminosity range of the highest-luminosity bin are 1.2, 1.5, and 1.1 for our study, GeV gamma-ray, and radio, respectively, and thus not so different. This could be attributed to that FSRQs with the highest hard X-ray luminosity does not necessarily have the highest gamma-ray or radio luminosity; it depends on the shape of spectral energy distribution. Higher number density of radio  FSRQs at high redshift indicates that a majority of radio luminous FSRQs are X-ray faint. X-ray emission of FSRQs is believed to be mainly from an external Compton component; a Compton scattering of low energy seed photons by high energy electrons. Such FSRQs may have a low radiation efficiency of the disk against a higher radiation efficiency of jet, and thus the seed photon flux is low, leading to a faint X-ray. Higher number density of gamma-ray FSRQs indicates that gamma-ray FSRQs have strong external Compton humps peaking at higher energies, and some of them are not detected by BAT.

From the Figure, the number of gamma-ray FSRQs could become comparable to that of hard X-ray FSRQs at higher redshift ($z\sim4$) due to the difference of the slope. This implies that high redshift FSRQs have steeper gamma-ray spectra and thus become leaked from {\it Fermi}/LAT detection. But, such high redshift objects could be detected by BAT. As a fact, 21~FSRQs in our sample does not locate within the error region of any FSRQs in the 4th LAT catalog (4FGL; \citet{LAT2019}). The fraction 40\% of no LAT counterpart in the BAT 105-month sample is remarkable regardless of a much larger number of LAT FSRQs. This  indicates the existence of hidden FSRQs with a faint gamma-ray luminosity but a bright X-ray luminosity. \citet{Paliya2017} also suggested that FSRQs not detected by Fermi/LAT has a lower inverse Compton peak energy than those detected by Fermi/LAT, based on multi-wavelength data.

\subsection{Origin of the Cosmic MeV Gamma-ray Background Radiation}

We found that FSRQs can not be the main contributor to the cosmic MeV gamma-ray background radiation. It may be made of other populations such as non-thermal coronal emission from radio-quiet AGNs \citep[e.g.,][]{Inoue2008, Inoue2019} and dark matter particles \citep{Olive1985,Ahn2005_DM1,Ahn2005_DM2}. Recently detection of non-thermal coronal synchrotron emission from radio-quiet AGNs is reported \citep{Inoue2018}, which suggests the existence of non-thermal electrons in the AGN coronae.

In order to probe the origin of the MeV background radiation, future gamma-ray observations of the entire sky by such as {\it e-ASTROGAM} \citep{DeAngelis2017}, {\it AMEGO} \citep{McEnery2019}, and {\it GRAMS} \citep{Aramaki2020} are important. However, even if they achieve their expected sensitivities, it is hard to resolve the MeV gamma-ray background radiation \citep{Inoue2015}. It is suggested that anisotropy measurements may dissolve the origin of the MeV gamma-ray background radiation \citep{Inoue2013_CXB}. Future MeV gamma-ray anisotropy observations will also be important to understand the origin of the MeV gamma-ray background radiation.

\section{Conclusion}
In this paper, we revisit the X-ray luminosity function of FSRQs utilizing the latest {\it Swift}/BAT 105 month catalog \citep{Oh2018}. The number of samples doubled comparing to previous works (A09). However, most of the newly identified objects are in the medium luminosity and/or low redshift regions. Adopting the LDDE model, the FSRQ evolution is found not to show extremely positive evolution, as reported in the previous works (A09).

By assuming spectral shape in the X-ray and gamma-ray bands, we also investigated the contribution of FSRQs to the cosmic MeV gamma-ray background radiation, where the origin is still uncertain. Utilizing our new FSRQ XLF, we found that FSRQs can explain $\sim3$\% of the cosmic MeV gamma-ray background radiation around 1~MeV. This suggests that other class objects are required to explain the cosmic MeV gamma-ray background radiation, such as non-thermal emission from Seyferts \citep{Inoue2008, Inoue2019} or annihilation of dark matter particles \citep{Olive1985, Ahn2005_DM1, Ahn2005_DM2}.

\acknowledgements
 YI is supported by JSPS KAKENHI Grant Number JP16K13813, JP18H05458, JP19K14772, program of Leading Initiative for Excellent Young Researchers, MEXT, Japan, and RIKEN iTHEMS Program. 

\bibliography{references}

\begin{thebibliography}{}
\expandafter\ifx\csname natexlab\endcsname\relax\def\natexlab#1{#1}\fi
\providecommand{\url}[1]{\href{#1}{#1}}

\bibitem[{{Ackermann} {et~al.}(2015){Ackermann}, {Ajello}, {Albert}, {Atwood},
  {Baldini}, {Ballet}, {Barbiellini}, {Bastieri}, {Bechtol}, {Bellazzini},
  {Bissaldi}, {Blandford}, {Bloom}, {Bottacini}, {Brandt}, {Bregeon}, {Bruel},
  {Buehler}, {Buson}, {Caliandro}, {Cameron}, {Caragiulo}, {Caraveo},
  {Cavazzuti}, {Cecchi}, {Charles}, {Chekhtman}, {Chiang}, {Chiaro}, {Ciprini},
  {Claus}, {Cohen-Tanugi}, {Conrad}, {Cuoco}, {Cutini}, {D'Ammando}, {de
  Angelis}, {de Palma}, {Dermer}, {Digel}, {Silva}, {Drell}, {Favuzzi},
  {Ferrara}, {Focke}, {Franckowiak}, {Fukazawa}, {Funk}, {Fusco}, {Gargano},
  {Gasparrini}, {Germani}, {Giglietto}, {Giommi}, {Giordano}, {Giroletti},
  {Godfrey}, {Gomez-Vargas}, {Grenier}, {Guiriec}, {Gustafsson}, {Hadasch},
  {Hayashi}, {Hays}, {Hewitt}, {Ippoliti}, {Jogler}, {J{\'o}hannesson},
  {Johnson}, {Johnson}, {Kamae}, {Kataoka}, {Kn{\"o}dlseder}, {Kuss},
  {Larsson}, {Latronico}, {Li}, {Li}, {Longo}, {Loparco}, {Lott}, {Lovellette},
  {Lubrano}, {Madejski}, {Manfreda}, {Massaro}, {Mayer}, {Mazziotta},
  {McEnery}, {Michelson}, {Mitthumsiri}, {Mizuno}, {Moiseev}, {Monzani},
  {Morselli}, {Moskalenko}, {Murgia}, {Nemmen}, {Nuss}, {Ohsugi}, {Omodei},
  {Orlando}, {Ormes}, {Paneque}, {Panetta}, {Perkins}, {Pesce-Rollins},
  {Piron}, {Pivato}, {Porter}, {Rain{\`o}}, {Rando}, {Razzano}, {Razzaque},
  {Reimer}, {Reimer}, {Reposeur}, {Ritz}, {Romani}, {S{\'a}nchez-Conde},
  {Schaal}, {Schulz}, {Sgr{\`o}}, {Siskind}, {Spandre}, {Spinelli}, {Strong},
  {Suson}, {Takahashi}, {Thayer}, {Thayer}, {Tibaldo}, {Tinivella}, {Torres},
  {Tosti}, {Troja}, {Uchiyama}, {Vianello}, {Werner}, {Winer}, {Wood}, {Wood},
  {Zaharijas}, \& {Zimmer}}]{Ackermann2015}
{Ackermann}, M., {Ajello}, M., {Albert}, A., {et~al.} 2015, \apj, 799, 86

\bibitem[{{Ahn} \& {Komatsu}(2005{\natexlab{a}})}]{Ahn2005_DM1}
{Ahn}, K., \& {Komatsu}, E. 2005{\natexlab{a}}, \prd, 71, 021303

\bibitem[{{Ahn} \& {Komatsu}(2005{\natexlab{b}})}]{Ahn2005_DM2}
---. 2005{\natexlab{b}}, \prd, 72, 061301

\bibitem[{{Ajello} {et~al.}(2008){Ajello}, {Greiner}, {Sato}, {Willis},
  {Kanbach}, {Strong}, {Diehl}, {Hasinger}, {Gehrels}, {Markwardt}, \&
  {Tueller}}]{Ajello2008}
{Ajello}, M., {Greiner}, J., {Sato}, G., {et~al.} 2008, \apj, 689, 666

\bibitem[{{Ajello} {et~al.}(2009){Ajello}, {Costamante}, {Sambruna}, {Gehrels},
  {Chiang}, {Rau}, {Escala}, {Greiner}, {Tueller}, {Wall}, \&
  {Mushotzky}}]{Ajello2009}
{Ajello}, M., {Costamante}, L., {Sambruna}, R.~M., {et~al.} 2009, \apj, 699,
  603

\bibitem[{{Ajello} {et~al.}(2012){Ajello}, {Shaw}, {Romani}, {Dermer},
  {Costamante}, {King}, {Max-Moerbeck}, {Readhead}, {Reimer}, {Richards}, \&
  {Stevenson}}]{Ajello2012}
{Ajello}, M., {Shaw}, M.~S., {Romani}, R.~W., {et~al.} 2012, \apj, 751, 108

\bibitem[{{Ajello} {et~al.}(2015){Ajello}, {Gasparrini}, {S{\'a}nchez-Conde},
  {Zaharijas}, {Gustafsson}, {Cohen-Tanugi}, {Dermer}, {Inoue}, {Hartmann},
  {Ackermann}, {Bechtol}, {Franckowiak}, {Reimer}, {Romani}, \&
  {Strong}}]{Ajello2015}
{Ajello}, M., {Gasparrini}, D., {S{\'a}nchez-Conde}, M., {et~al.} 2015, \apjl,
  800, L27

\bibitem[{{Aramaki} {et~al.}(2020){Aramaki}, {Adrian}, {Karagiorgi}, \&
  {Odaka}}]{Aramaki2020}
{Aramaki}, T., {Adrian}, P. O.~H., {Karagiorgi}, G., \& {Odaka}, H. 2020,
  Astroparticle Physics, 114, 107

\bibitem[{{Atwood} {et~al.}(2009){Atwood}, {Abdo}, {Ackermann}, {Althouse},
  {Anderson}, {Axelsson}, {Baldini}, {Ballet}, {Band}, {Barbiellini}, \&
  et~al.}]{Atwood2009}
{Atwood}, W.~B., {Abdo}, A.~A., {Ackermann}, M., {et~al.} 2009, \apj, 697, 1071

\bibitem[{{Ba{\~n}ados} {et~al.}(2018){Ba{\~n}ados}, {Venemans},
  {Mazzucchelli}, {Farina}, {Walter}, {Wang}, {Decarli}, {Stern}, {Fan},
  {Davies}, {Hennawi}, {Simcoe}, {Turner}, {Rix}, {Yang}, {Kelson}, {Rudie}, \&
  {Winters}}]{Banados2018}
{Ba{\~n}ados}, E., {Venemans}, B.~P., {Mazzucchelli}, C., {et~al.} 2018, \nat,
  553, 473

\bibitem[{{Blandford} \& {Znajek}(1977)}]{Blandford1977}
{Blandford}, R.~D., \& {Znajek}, R.~L. 1977, \mnras, 179, 433

\bibitem[{{Blom} {et~al.}(1995){Blom}, {Bennett}, {Bloemen}, {Collmar},
  {Diehl}, {Hermsen}, {Iyudin}, {Schoenfelder}, {Stacy}, {Steinle}, {Williams},
  \& {Winkler}}]{Blom1995}
{Blom}, J.~J., {Bennett}, K., {Bloemen}, H., {et~al.} 1995, \aap, 298, L33

\bibitem[{{Churazov} {et~al.}(2007){Churazov}, {Sunyaev}, {Revnivtsev},
  {Sazonov}, {Molkov}, {Grebenev}, {Winkler}, {Parmar}, {Bazzano}, {Falanga},
  {Gros}, {Lebrun}, {Natalucci}, {Ubertini}, {Roques}, {Bouchet}, {Jourdain},
  {Kn{\"o}dlseder}, {Diehl}, {Budtz-Jorgensen}, {Brandt}, {Lund},
  {Westergaard}, {Neronov}, {T{\"u}rler}, {Chernyakova}, {Walter}, {Produit},
  {Mowlavi}, {Mas-Hesse}, {Domingo}, {Gehrels}, {Kuulkers}, {Kretschmar}, \&
  {Schmidt}}]{churazov2007}
{Churazov}, E., {Sunyaev}, R., {Revnivtsev}, M., {et~al.} 2007, \aap, 467, 529

\bibitem[{{De Angelis} {et~al.}(2017){De Angelis}, {Tatischeff}, {Tavani},
  {Oberlack}, {Grenier}, {Hanlon}, {Walter}, {Argan}, {von Ballmoos},
  {Bulgarelli}, {Donnarumma}, {Hernanz}, {Kuvvetli}, {Pearce}, {Zdziarski},
  {Aboudan}, {Ajello}, {Ambrosi}, {Bernard}, {Bernardini}, {Bonvicini},
  {Brogna}, {Branchesi}, {Budtz-Jorgensen}, {Bykov}, {Campana}, {Cardillo},
  {Coppi}, {De Martino}, {Diehl}, {Doro}, {Fioretti}, {Funk}, {Ghisellini},
  {Grove}, {Hamadache}, {Hartmann}, {Hayashida}, {Isern}, {Kanbach}, {Kiener},
  {Kn{\"o}dlseder}, {Labanti}, {Laurent}, {Limousin}, {Longo}, {Mannheim},
  {Marisaldi}, {Martinez}, {Mazziotta}, {McEnery}, {Mereghetti}, {Minervini},
  {Moiseev}, {Morselli}, {Nakazawa}, {Orleanski}, {Paredes}, {Patricelli},
  {Peyr{\'e}}, {Piano}, {Pohl}, {Ramarijaona}, {Rando}, {Reichardt},
  {Roncadelli}, {Silva}, {Tavecchio}, {Thompson}, {Turolla}, {Ulyanov},
  {Vacchi}, {Wu}, \& {Zoglauer}}]{DeAngelis2017}
{De Angelis}, A., {Tatischeff}, V., {Tavani}, M., {et~al.} 2017, Experimental
  Astronomy, 44, 25

\bibitem[{{Fossati} {et~al.}(1998){Fossati}, {Maraschi}, {Celotti}, {Comastri},
  \& {Ghisellini}}]{Fossati1998}
{Fossati}, G., {Maraschi}, L., {Celotti}, A., {Comastri}, A., \& {Ghisellini},
  G. 1998, \mnras, 299, 433

\bibitem[{{Fukada} {et~al.}(1975){Fukada}, {Hayakawa}, {Kasahara}, {Makino},
  {Tanaka}, \& {Sreekantan}}]{Fukada1975}
{Fukada}, Y., {Hayakawa}, S., {Kasahara}, I., {et~al.} 1975, \nat, 254, 398

\bibitem[{{Gehrels} {et~al.}(2004){Gehrels}, {Chincarini}, {Giommi}, {Mason},
  {Nousek}, {Wells}, {White}, {Barthelmy}, {Burrows}, {Cominsky}, {Hurley},
  {Marshall}, {M{\'e}sz{\'a}ros}, {Roming}, {Angelini}, {Barbier}, {Belloni},
  {Campana}, {Caraveo}, {Chester}, {Citterio}, {Cline}, {Cropper}, {Cummings},
  {Dean}, {Feigelson}, {Fenimore}, {Frail}, {Fruchter}, {Garmire}, {Gendreau},
  {Ghisellini}, {Greiner}, {Hill}, {Hunsberger}, {Krimm}, {Kulkarni}, {Kumar},
  {Lebrun}, {Lloyd-Ronning}, {Markwardt}, {Mattson}, {Mushotzky}, {Norris},
  {Osborne}, {Paczynski}, {Palmer}, {Park}, {Parsons}, {Paul}, {Rees},
  {Reynolds}, {Rhoads}, {Sasseen}, {Schaefer}, {Short}, {Smale}, {Smith},
  {Stella}, {Tagliaferri}, {Takahashi}, {Tashiro}, {Townsley}, {Tueller},
  {Turner}, {Vietri}, {Voges}, {Ward}, {Willingale}, {Zerbi}, \&
  {Zhang}}]{Gehrels2004}
{Gehrels}, N., {Chincarini}, G., {Giommi}, P., {et~al.} 2004, \apj, 611, 1005

\bibitem[{{Gendreau} {et~al.}(1995){Gendreau}, {Mushotzky}, {Fabian}, {Holt},
  {Kii}, {Serlemitsos}, {Ogasaka}, {Tanaka}, {Bautz}, {Fukazawa}, {Ishisaki},
  {Kohmura}, {Makishima}, {Tashiro}, {Tsusaka}, {Kunieda}, {Ricker}, \&
  {Vanderspek}}]{Gendreau1995}
{Gendreau}, K.~C., {Mushotzky}, R., {Fabian}, A.~C., {et~al.} 1995, \pasj, 47,
  L5

\bibitem[{{Ghisellini} {et~al.}(2017){Ghisellini}, {Righi}, {Costamante}, \&
  {Tavecchio}}]{Ghisellini2017}
{Ghisellini}, G., {Righi}, C., {Costamante}, L., \& {Tavecchio}, F. 2017,
  \mnras, 469, 255

\bibitem[{{Ghisellini} {et~al.}(2014){Ghisellini}, {Tavecchio}, {Maraschi},
  {Celotti}, \& {Sbarrato}}]{Ghisellini2014}
{Ghisellini}, G., {Tavecchio}, F., {Maraschi}, L., {Celotti}, A., \&
  {Sbarrato}, T. 2014, \nat, 515, 376

\bibitem[{{Gilli} {et~al.}(2007){Gilli}, {Comastri}, \& {Hasinger}}]{Gilli2007}
{Gilli}, R., {Comastri}, A., \& {Hasinger}, G. 2007, \aap, 463, 79

\bibitem[{{Gruber} {et~al.}(1999){Gruber}, {Matteson}, {Peterson}, \&
  {Jung}}]{Gruber1999}
{Gruber}, D.~E., {Matteson}, J.~L., {Peterson}, L.~E., \& {Jung}, G.~V. 1999,
  \apj, 520, 124

\bibitem[{{Inoue} \& {Doi}(2018)}]{Inoue2018}
{Inoue}, Y., \& {Doi}, A. 2018, \apj, 869, 114

\bibitem[{{Inoue} {et~al.}(2017){Inoue}, {Doi}, {Tanaka}, {Sikora}, \&
  {Madejski}}]{Inoue2017}
{Inoue}, Y., {Doi}, A., {Tanaka}, Y.~T., {Sikora}, M., \& {Madejski}, G.~M.
  2017, \apj, 840, 46

\bibitem[{{Inoue} {et~al.}(2019){Inoue}, {Khangulyan}, {Inoue}, \&
  {Doi}}]{Inoue2019}
{Inoue}, Y., {Khangulyan}, D., {Inoue}, S., \& {Doi}, A. 2019, \apj, 880, 40

\bibitem[{{Inoue} {et~al.}(2013){Inoue}, {Murase}, {Madejski}, \&
  {Uchiyama}}]{Inoue2013_CXB}
{Inoue}, Y., {Murase}, K., {Madejski}, G.~M., \& {Uchiyama}, Y. 2013, \apj,
  776, 33

\bibitem[{{Inoue} {et~al.}(2015){Inoue}, {Tanaka}, {Odaka}, {Takada},
  {Ichinohe}, {Saito}, {Takeda}, \& {Takahashi}}]{Inoue2015}
{Inoue}, Y., {Tanaka}, Y.~T., {Odaka}, H., {et~al.} 2015, \pasj, 67, 76

\bibitem[{{Inoue} \& {Totani}(2009)}]{Inoue2009}
{Inoue}, Y., \& {Totani}, T. 2009, \apj, 702, 523

\bibitem[{{Inoue} {et~al.}(2008){Inoue}, {Totani}, \& {Ueda}}]{Inoue2008}
{Inoue}, Y., {Totani}, T., \& {Ueda}, Y. 2008, \apj, 672, L5

\bibitem[{{Kinzer} {et~al.}(1997){Kinzer}, {Jung}, {Gruber}, {Matteson},
  {Peterson}, \& {L.~E.}}]{Kinzer1997}
{Kinzer}, R.~L., {Jung}, G.~V., {Gruber}, D.~E., {et~al.} 1997, \apj, 475, 361

\bibitem[{{Kubo} {et~al.}(1998){Kubo}, {Takahashi}, {Madejski}, {Tashiro},
  {Makino}, {Inoue}, \& {Takahara}}]{Kubo1998}
{Kubo}, H., {Takahashi}, T., {Madejski}, G., {et~al.} 1998, \apj, 504, 693

\bibitem[{{La Franca} \& {Cristiani}(1997)}]{laFranca1997}
{La Franca}, F., \& {Cristiani}, S. 1997, \aj, 113, 1517

\bibitem[{{Mao} {et~al.}(2017){Mao}, {Urry}, {Marchesini}, {Landoni},
  {Massaro}, \& {Ajello}}]{Mao2017}
{Mao}, P., {Urry}, C.~M., {Marchesini}, E., {et~al.} 2017, \apj, 842, 87

\bibitem[{{Massaro} {et~al.}(2015){Massaro}, {Maselli}, {Leto}, {Marchegiani},
  {Perri}, {Giommi}, \& {Piranomonte}}]{Massaro2015}
{Massaro}, E., {Maselli}, A., {Leto}, C., {et~al.} 2015, \apss, 357, 75

\bibitem[{{Matsuoka} {et~al.}(2018){Matsuoka}, {Strauss}, {Kashikawa}, {Onoue},
  {Iwasawa}, {Tang}, {Lee}, {Imanishi}, {Nagao}, {Akiyama}, {Asami}, {Bosch},
  {Furusawa}, {Goto}, {Gunn}, {Harikane}, {Ikeda}, {Izumi}, {Kawaguchi},
  {Kato}, {Kikuta}, {Kohno}, {Komiyama}, {Lupton}, {Minezaki}, {Miyazaki},
  {Murayama}, {Niida}, {Nishizawa}, {Noboriguchi}, {Oguri}, {Ono}, {Ouchi},
  {Price}, {Sameshima}, {Schulze}, {Shirakata}, {Silverman}, {Sugiyama},
  {Tait}, {Takada}, {Takata}, {Tanaka}, {Toba}, {Utsumi}, {Wang}, \&
  {Yamashita}}]{Matsuoka2018}
{Matsuoka}, Y., {Strauss}, M.~A., {Kashikawa}, N., {et~al.} 2018, \apj, 869,
  150

\bibitem[{{McEnery} {et~al.}(2019){McEnery}, {Abel Barrio}, {Agudo}, {Ajello},
  {{\'A}lvarez}, {Ansoldi}, {Anton}, {Auricchio}, {Stephen}, {Baldini},
  {Bambi}, {Baring}, {Barres}, {Bastieri}, {Beacom}, {Beckmann}, {Bednarek},
  {Bernard}, {Bissaldi}, {Bloser}, {Blumer}, {Boettcher}, {Boggs},
  {Bolotnikov}, {Bottacini}, {Bozhilov}, {Bozzo}, {Briggs}, {Buckley}, {Buson},
  {Campana}, {Caputo}, {Cardillo}, {Caroli}, {Castro}, {Cenko}, {Charles},
  {Chen}, {Cheung}, {Ciprini}, {Coppi}, {Curado da Silva}, {Cutini}, {D'Ammand
  o}, {De Angelis}, {De Becker}, {De Nolfo}, {Del Sordo}, {Di Mauro}, {Di
  Venere}, {Dietrich}, {Digel}, {Dominguez}, {Doro}, {Ferrara}, {Fields},
  {Finke}, {Foffano}, {Fryer}, {Fukazawa}, {Funk}, {Gasparrini}, {Gelfand},
  {Georganopoulos}, {Giordano}, {Giuliani}, {Gouiffes}, {Grefenstette},
  {Grenier}, {Griffin}, {Grove}, {Guiriec}, {Harding}, {Harding}, {Hartmann},
  {Hays}, {Hernanz}, {Hewitt}, {Holder}, {Hui}, {Inglis}, {Johnson}, {Jones},
  {Kanbach}, {Kargaltsev}, {Kaufmann}, {Kerr}, {Kierans}, {Kislat}, {Klimenko},
  {Knodlseder}, {Kocveski}, {Kopp}, {Krawczynsiki}, {Krizmanic}, {Kubo},
  {Kurahashi Neilson}, {Laurent}, {Lenain}, {Li}, {Lien}, {Linden}, {Lommler},
  {Longo}, {Lovellette}, {L{\'o}pez}, {Manousakis}, {Marcotulli}, {Marcowith},
  {Martinez}, {McConnell}, {Metcalfe}, {Meyer}, {Meyer}, {Mignani}, {Mitchell},
  {Mizuno}, {Moiseev}, {Morcuende}, {Moskalenko}, {Moss}, {Nakazawa},
  {Mazziotta}, {Oberlack}, {Ohno}, {Oikonomou}, {Ojha}, {Omodei}, {Orlando},
  {Otte}, {Paliya}, {Parker}, {Patricelli}, {Perkins}, {Petropoulou},
  {Pittori}, {Pohl}, {Porter}, {Prandini}, {Prescod-Weinstein}, {Racusin},
  {Rand o}, {Rani}, {Rib{\'o}}, {Rodi}, {Sanchez-Conde}, {Saz Parkinson},
  {Schirato}, {Shawhan}, {Shrader}, {Smith}, {Smith}, {Stamerra}, {Stawarz},
  {Strong}, {Stumke}, {Tajima}, {Takahashi}, {Tanaka}, {Tatischeff}, {The},
  {Thompson}, {Tibaldo}, {Tomsick}, {Uhm}, {Venters}, {Vestrand}, {Vianello},
  {Wadiasingh}, {Walter}, {Wang}, {Williams}, {Wilson-Hodge}, {Wood}, {Woolf},
  {Wulf}, {Younes}, {Zampieri}, {Zane}, {Zhang}, {Zhang}, {Zimmer}, {Zoglauer},
  \& {van der Horst}}]{McEnery2019}
{McEnery}, J., {Abel Barrio}, J., {Agudo}, I., {et~al.} 2019, arXiv e-prints,
  arXiv:1907.07558

\bibitem[{{Miyaji} {et~al.}(2001){Miyaji}, {Hasinger}, \&
  {Schmidt}}]{Miyaji2001}
{Miyaji}, T., {Hasinger}, G., \& {Schmidt}, M. 2001, \aap, 369, 49

\bibitem[{{Narumoto} \& {Totani}(2006)}]{Narumoto2006}
{Narumoto}, T., \& {Totani}, T. 2006, \apj, 643, 81

\bibitem[{{Oh} {et~al.}(2018){Oh}, {Koss}, {Markwardt}, {Schawinski},
  {Baumgartner}, {Barthelmy}, {Cenko}, {Gehrels}, {Mushotzky}, {Petulante},
  {Ricci}, {Lien}, \& {Trakhtenbrot}}]{Oh2018}
{Oh}, K., {Koss}, M., {Markwardt}, C.~B., {et~al.} 2018, \apjs, 235, 4

\bibitem[{{Olive} \& {Silk}(1985)}]{Olive1985}
{Olive}, K.~A., \& {Silk}, J. 1985, Physical Review Letters, 55, 2362

\bibitem[{{Padovani} {et~al.}(1993){Padovani}, {Ghisellini}, {Fabian}, \&
  {Celotti}}]{Padovani1993}
{Padovani}, P., {Ghisellini}, G., {Fabian}, A.~C., \& {Celotti}, A. 1993,
  \mnras, 260, L21

\bibitem[{{Paliya} {et~al.}(2017){Paliya}, {Marcotulli}, {Ajello}, {Joshi},
  {Sahayanathan}, {Rao}, \& {Hartmann}}]{Paliya2017}
{Paliya}, V.~S., {Marcotulli}, L., {Ajello}, M., {et~al.} 2017, \apj, 851, 33

\bibitem[{{Revnivtsev} {et~al.}(2003){Revnivtsev}, {Gilfanov}, {Sunyaev},
  {Jahoda}, \& {Markwardt}}]{Revnivtsev2003}
{Revnivtsev}, M., {Gilfanov}, M., {Sunyaev}, R., {Jahoda}, K., \& {Markwardt},
  C. 2003, \aap, 411, 329

\bibitem[{{Schmidt}(1963)}]{Schmidt1963}
{Schmidt}, M. 1963, \nat, 197, 1040

\bibitem[{{Shaw} {et~al.}(2013){Shaw}, {Filippenko}, {Romani}, {Cenko}, \&
  {Li}}]{Shaw2013}
{Shaw}, M.~S., {Filippenko}, A.~V., {Romani}, R.~W., {Cenko}, S.~B., \& {Li},
  W. 2013, \aj, 146, 127

\bibitem[{{Sikora} {et~al.}(2007){Sikora}, {Stawarz}, \& {Lasota}}]{Sikora2007}
{Sikora}, M., {Stawarz}, {\L}., \& {Lasota}, J.-P. 2007, \apj, 658, 815

\bibitem[{{The Fermi-LAT collaboration}(2019)}]{LAT2019}
{The Fermi-LAT collaboration}. 2019, arXiv e-prints, arXiv:1902.10045

\bibitem[{{Ueda} {et~al.}(2014){Ueda}, {Akiyama}, {Hasinger}, {Miyaji}, \&
  {Watson}}]{Ueda2014}
{Ueda}, Y., {Akiyama}, M., {Hasinger}, G., {Miyaji}, T., \& {Watson}, M.~G.
  2014, \apj, 786, 104

\bibitem[{{Urry} \& {Padovani}(1995)}]{Urry1995}
{Urry}, C.~M., \& {Padovani}, P. 1995, Publications of the Astronomical Society
  of the Pacific, 107, 803

\bibitem[{{Watanabe} {et~al.}(1997){Watanabe}, {Hartmann}, {Leising}, {The},
  {Share}, \& {Kinzer}}]{Watanabe1997}
{Watanabe}, K., {Hartmann}, D.~H., {Leising}, M.~D., {et~al.} 1997, in American
  Institute of Physics Conference Series, Vol. 410, Proceedings of the Fourth
  Compton Symposium, ed. C.~D. {Dermer}, M.~S. {Strickman}, \& J.~D. {Kurfess},
  1223--1227

\bibitem[{{Weidenspointner} {et~al.}(2000){Weidenspointner}, {Varendorff},
  {Kappadath}, {Bennett}, {Bloemen}, {Diehl}, {Hermsen}, {Lichti}, {Ryan}, \&
  {Sch{\"o}nfelder}}]{Weidenspointner2000}
{Weidenspointner}, G., {Varendorff}, M., {Kappadath}, S.~C., {et~al.} 2000, in
  American Institute of Physics Conference Series, Vol. 510, American Institute
  of Physics Conference Series, ed. M.~L. {McConnell} \& J.~M. {Ryan}, 467--470

\end{thebibliography}

\newpage

\startlongtable
\begin{deluxetable*}{lllcccc}
\tablecaption{Our sample FSRQs from BAT 105 month catalog}
%\hspace{-1.8cm}
\tablehead{
\colhead{SWIFT~name} & \colhead{BZCAT~name} & \colhead{redshift} & \colhead{RA} & \colhead{DEC} & \colhead{Flux$^a$} & \colhead{Photon Index$^b$} \\
}
\startdata
J0010.5+1057$^{\ast}$ & 5BZQJ0010+1058 & 0.089 & 2.62917 & 10.97489 & 30.34 & 1.82 \\ 
 J0017.1+8134$^{\ast}$ & 5BZQJ0017+8135 & 3.366 & 4.28525 & 81.58561 & 11.39 & 2.42 \\ 
 J0144.8-2754 & 5BZQJ0145-2733 & 1.155 & 26.26413 & -27.55953 & 10.72 & 1.43 \\ 
 J0225.0+1847$^{\ast}$ & 5BZQJ0225+1846 & 2.69 & 36.26946 & 18.78022 & 31.37 & 1.73 \\ 
 J0233.8+0243 & 5BZQJ0233+0229 & 0.321 & 38.45496 & 2.49031 & 7.02 & 2.69 \\ 
 J0311.8-7653$^{\ast}$ & 5BZQJ0311-7651 & 0.223 & 47.98021 & -76.86414 & 10.47 & 1.98 \\ 
 J0336.6+3217$^{\ast}$ & 5BZQJ0336+3218 & 1.258 & 54.12546 & 32.30814 & 44.17 & 1.67 \\ 
 J0404.0-3604 & 5BZQJ0403-3605 & 1.417 & 60.97396 & -36.08386 & 10.65 & 1.91 \\ 
 J0405.5-1307 & 5BZQJ0405-1308 & 0.57 & 61.39167 & -13.13714 & 11.03 & 1.78 \\ 
 J0525.1-2339 & 5BZQJ0525-2338 & 3.1 & 81.27713 & -23.63633 & 13.11 & 1.55 \\ 
 J0525.3-4600$^{\ast}$ & 5BZQJ0525-4557 & 1.479 & 81.38083 & -45.96519 & 15.51 & 1.37 \\ 
 J0539.9-2839$^{\ast}$ & 5BZQJ0539-2839 & 3.104 & 84.97617 & -28.66553 & 29.01 & 1.33 \\ 
 J0547.1-6427 & 5BZQJ0546-6415 & 0.323 & 86.67433 & -64.25608 & 8.21 & 1.95 \\ 
 J0623.3-6438 & 5BZQJ0623-6436 & 0.128 & 95.78204 & -64.60578 & 11.64 & 1.98 \\ 
 J0635.8-7514$^{\ast}$ & 5BZQJ0635-7516 & 0.651 & 98.94379 & -75.27133 & 16.52 & 2 \\ 
 J0746.3+2548$^{\ast}$ & 5BZQJ0746+2549 & 2.979 & 116.60779 & 25.81725 & 36.01 & 1.43 \\ 
 J0805.2+6145$^{\ast}$ & 5BZQJ0805+6144 & 3.033 & 121.32562 & 61.74 & 17.53 & 1.35 \\ 
 J0841.4+7052$^{\ast}$ & 5BZQJ0841+7053 & 2.172 & 130.3515 & 70.89508 & 69.81 & 1.7 \\ 
 J0842.0+4021 & 5BZQJ0842+4018 & 0.151 & 130.51558 & 40.30875 & 6.93 & 2.41 \\ 
 J1044.8+8091 & 5BZQJ1044+8054 & 1.26 & 161.09608 & 80.91094 & 11.73 & 1.67 \\ 
 J1130.1-1447$^{\ast}$ & 5BZQJ1130-1449 & 1.184 & 172.52938 & -14.82428 & 28.74 & 1.88 \\ 
 J1153.0+3311 & 5BZQJ1152+3307 & 1.397 & 178.21629 & 33.12189 & 10.08 & 1.83 \\ 
 J1153.6+4931 & 5BZQJ1153+4931 & 0.334 & 178.35196 & 49.51911 & 12.78 & 1.83 \\ 
 J1159.7+2923 & 5BZQJ1159+2914 & 0.724 & 179.88262 & 29.2455 & 7.49 & 1.84 \\ 
 J1222.4+0414 & 5BZQJ1222+0413 & 0.965 & 185.59392 & 4.22103 & 36.22 & 1.45 \\ 
 J1224.9+2122$^{\ast}$ & 5BZQJ1224+2122 & 0.432 & 186.22692 & 21.37956 & 24.5 & 1.7 \\ 
 J1229.1+0202$^{\ast}$ & 5BZQJ1229+0203 & 0.158 & 187.27792 & 2.05239 & 421.57 & 1.75 \\ 
 J1256.2-0551$^{\ast}$ & 5BZQJ1256-0547 & 0.536 & 194.04654 & -5.78931 & 38.82 & 1.32 \\ 
 J1305.4-1034 & 5BZQJ1305-1033 & 0.278 & 196.38758 & -10.55539 & 13.72 & 1.7 \\ 
 J1331.6-0504 & 5BZQJ1332-0509 & 2.15 & 203.01858 & -5.16203 & 15.5 & 1.51 \\ 
 J1337.7-1253 & 5BZQJ1337-1257 & 0.539 & 204.41575 & -12.95686 & 13.21 & 2.19 \\ 
 J1357.0+1929 & 5BZQJ1357+1919 & 0.72 & 209.2685 & 19.31872 & 8.67 & 2.02 \\ 
 J1430.6+4211 & 5BZQJ1430+4204 & 4.715 & 217.59892 & 42.07681 & 9.9 & 1.56 \\ 
 J1512.8-0906$^{\ast}$ & 5BZQJ1512-0905 & 0.36 & 228.21054 & -9.09994 & 66.8 & 1.32 \\ 
 J1557.8-7913 & 5BZQJ1556-7914 & 0.15 & 239.24529 & -79.23453 & 14.98 & 2.31 \\ 
 J1625.9+4349 & 5BZQJ1625+4347 & 1.048 & 246.47213 & 43.78717 & 12.13 & 2.04 \\ 
 J1643.1+3951 & 5BZQJ1642+3948 & 0.592 & 250.74504 & 39.81028 & 20.71 & 1.17 \\ 
 J1658.5+0518 & 5BZQJ1658+0515 & 0.879 & 254.63938 & 5.25456 & 12.75 & 1.79 \\ 
 J1848.5+6704 & 5BZQJ1849+6705 & 0.657 & 282.31696 & 67.09492 & 6.39 & 2.72 \\ 
 J1924.9-2918 & 5BZQJ1924-2914 & 0.352 & 291.21275 & -29.24169 & 16.22 & 2.04 \\ 
 J1928.0+7356 & 5BZQJ1927+7358 & 0.302 & 291.95208 & 73.96711 & 11.04 & 2.5 \\ 
 J2011.5-1544 & 5BZQJ2011-1546 & 1.18 & 302.81546 & -15.77786 & 12.6 & 2.41 \\ 
 J2129.1-1538$^{\ast}$ & 5BZQJ2129-1538 & 3.268 & 322.30075 & -15.64472 & 20.05 & 1.79 \\ 
 J2148.0+0657 & 5BZQJ2148+0657 & 0.99 & 327.02275 & 6.96072 & 17.37 & 1.9 \\ 
 J2148.4-7557 & 5BZQJ2147-7536 & 1.139 & 326.80304 & -75.60367 & 14.59 & 1.41 \\ 
 J2152.0-3030$^{\ast}$ & 5BZQJ2151-3027 & 2.345 & 327.98133 & -30.46492 & 89.3 & 1.61 \\ 
 J2203.0+3146 & 5BZQJ2203+3145 & 0.295 & 330.81242 & 31.76064 & 15.56 & 1.92 \\ 
 J2211.7+1843 & 5BZQJ2211+1841 & 0.07 & 332.97454 & 18.69719 & 15.72 & 1.88 \\ 
 J2229.7-0831$^{\ast}$ & 5BZQJ2229-0832 & 1.559 & 337.417 & -8.54844 & 17.08 & 1.46 \\ 
 J2232.5+1141$^{\ast}$ & 5BZQJ2232+1143 & 1.037 & 338.15171 & 11.73081 & 30.05 & 1.49 \\ 
 J2251.9+2215$^{\ast}$ & 5BZQJ2251+2217 & 3.668 & 342.97292 & 22.29369 & 9.61 & 2.36 \\ 
 J2253.9+1608$^{\ast}$ & 5BZQJ2253+1608 & 0.859 & 343.49063 & 16.14822 & 158.36 & 1.5 \\ 
 J2327.5+0938$^{\ast}$ & 5BZQJ2327+0940 & 1.843 & 351.88992 & 9.66931 & 29.73 & 1.4 \\ 
\enddata
\label{sample}
$a$: BAT X-ray flux in unit of $10^{-12}$ erg cm$^{-2}$ s$^{-1}$ (14--195 keV).\\
$b$: BAT X-ray photon index.\\
$\ast$: BAT FSRQ listed in A09.\\
\end{deluxetable*}

\end{document}